\documentclass{article}
\pdfoutput=1
\usepackage{graphicx} 
\usepackage{authblk}
\usepackage[a4paper, total={6in, 8in}]{geometry}
\RequirePackage[T1]{fontenc}
\RequirePackage[utf8]{inputenc}
\RequirePackage{calc}
\RequirePackage{indentfirst}
\RequirePackage{fancyhdr}
\RequirePackage{graphicx,epstopdf}
\RequirePackage{lastpage}
\RequirePackage{ifthen}
\RequirePackage{float}
\RequirePackage{amsmath}
\RequirePackage{amssymb} 
\RequirePackage[right]{lineno}
\RequirePackage{setspace}
\RequirePackage{enumitem}
\RequirePackage{mathpazo}
\RequirePackage{booktabs} 
\RequirePackage{titlesec}
\RequirePackage{etoolbox} 
\RequirePackage{tabto} 
\RequirePackage{xcolor, colortbl} 
\RequirePackage{soul} 

\RequirePackage{multirow}
\RequirePackage{microtype} 
\RequirePackage{tikz} 
\RequirePackage{changepage} 
\RequirePackage{attrib} 
\RequirePackage{upgreek} 
\RequirePackage{array} 
\RequirePackage{tabularx}

\title{Dynamic Bayesian Networks, Elicitation and Data Embedding
for Secure Environments}
\author[1]{Kieran Drury}
\author[1]{Jim Q. Smith}
\affil[1]{Department of Statistics, University of Warwick, Coventry CV4 7AL, UK}
\date{11 September 2024}

\begin{document}

\maketitle

\begin{abstract}
    Serious crime modelling typically needs to be undertaken securely behind a firewall where police knowledge and capabilities can remain undisclosed. Data informing an ongoing incident is often sparse, with a large proportion of relevant data only coming to light after the incident culminates or after police intervene - by which point it is too late to make use of the data to aid real-time decision making for the incident in question. Much of the data that \emph{is} available to police to support real-time decision making is highly confidential so cannot be shared with academics, and is therefore missing to them. In this paper, we describe the development of a formal protocol where a graphical model is used as a framework for securely translating a model designed by an academic team to a model for use by a police team. We then show, for the first time, how libraries of these models can be built and used for real-time decision support to circumvent the challenges of data missingness and tardiness seen in such a secure environment. The parallel development described by this protocol ensures that any sensitive information collected by police, and missing to academics, remains secured behind a firewall. The protocol nevertheless guides police so that they are able to combine the typically incomplete data streams that are open source with their more sensitive information in a formal and justifiable way. We illustrate the application of this protocol by describing how a new entry - a suspected vehicle attack - can be embedded into such a police library of criminal plots.
\end{abstract}


\section{Introduction}

Over the last 15 years - starting with dynamic Bayesian Networks (DBNs) - graphical
models have been developed to provide police with probabilistic predictive
models of various orchestrated criminal activities. Typically, these domains
are very fast moving and various components of critical informative data streams are often systematically missing. The missingness
is usually in no sense at random and often comes as a result of data relevant to an ongoing criminal plot becoming available only after the plot has either been carried out or prevented. Data that is useful for guiding police in the intervention of such plots often cannot be observed by police at the point where they must decide whether to intervene, usually appearing after this point in time in the form of court hearings and incident reports. These issues often require models to be
customised in subtle ways so that predictive algorithms are fit for purpose \cite{BunninSmith2021}. They also strongly benefit from police being able to match a current ongoing plot to a pre-existing embellished probability model from a previous incident which can be then adjusted to suit the ongoing plot as fit. This process is aided through the use of \textit{libraries} which we discuss in this paper as a mechanism to tackle this type of missingness of data that we see in this application.

Yet there exists an even bigger challenge. When performing this modelling task,
police must ensure that the types and extent of the various streams of data
available to them, and the precise algorithms they use to make inferences, are
not directly betrayed to a suspected criminal who could then use this
information for their own advantage. So any operationalisable decision
analysis based on these probabilistic graphical models typically needs to be
undertaken\emph{\ securely} behind a firewall where police knowledge and
capabilities are kept secret. Therefore, data that would otherwise inform police behind the firewall of the critical features of the underlying
process is not only systematically missing or disguised, but must be kept secret from - and is therefore
missing to - any remote team of specialist decision analysts and statistical
modellers supporting police in developing decision models. These academic specialist modellers are typically required for the development of advanced, bespoke models of the kind police will need for effective decision support. Typically, police only have limited in-house modelling resources. It is therefore of the highest importance to develop a co-creation protocol that enables the sharing of both academic expertise and police intelligence securely across the firewall.

Such protocols allow police to directly apply state-of-the-art inferential and elicitation methodologies to
systematically address and surmount these very particular challenges that they would otherwise struggle to overcome.
The protocols we report here - to develop co-created libraries of
customised graphical models - have now been successfully applied by
co-creation teams across a wide range of secure use cases. The co-creating
teams begin by building generic frameworks that \emph{describe and} \emph{%
categorise} incidents of crime. Policing agents are then trained by academic
teams to match an ongoing incident to a particular graphical model. The
structural information embedded in each graph enables the academic team to
guide police in building a library of embellishing probability models
around these frameworks which can then support police in frustrating crimes
that are planned to harm the general public. In this paper, we focus on those
police libraries where the appropriate choice of structural framework for
each category of unfolding criminal incident can be represented by the graph
of a Bayesian Network (BN), before then embellishing this into a full probability model by populating its conditional probability tables (CPTs).
The protocol we have been developing over the last ten years is reported here for the first time and provides a generic yet detailed methodology for this.

A central concept behind this protocol is the synergetic communication across \emph{parallel inferential code}, either side of a firewall between two teams. Thus, using open source data, an academic team of analysts
outside the firewall first elicit families of probability models consistent
with generic developments associated with particular
categories of crime. Through a co-creation scheme of sequential interactions
with a parallel team of police, the academic team then incrementally build
up a library of coded-up probability models - one for each category of crime
- each described by its own BN. Parallel coded models simultaneously
developed, one by a team of academics and another by police - enhanced by the shared
structure of BNs either side of the firewall - then provide the framework
for symbiotic embellishments in ways we describe below. Using the Bayesian
paradigm, these embellishments then enable police to predict the progress of
new incidents as these unfold - both when police simply observe the
incident and when police consider intervening in various ways. These predictions
can then be used to inform a decision support system designed to frustrate
the progress of the crime and mitigate its potential harm.

The co-creation protocol we describe here enables defenders to efficiently
and effectively communicate with professional decision support experts who
are not necessarily sufficiently security cleared to nevertheless guide the
development of an appropriate stochastic model. In particular, police, so
guided by an academic team, are able to maintain the security of sensitive
information to build up their own secure libraries of BNs.

Within the context of policing, it is clearly essential that the graphical
framework provides not only predictions of what might happen if police
simply watch the progress of a crime, but also what might happen if they intervene. In technical terminology, this requires the BN to be causal. In Section 2, we give a brief review of such causal challenges as these might
apply to this particular type of adversarial domain, and describe how a Bayesian decision analytic paradigm provides a
natural framework for such secured co-creation. Then, in Section 3, we
describe the types of libraries of graphical probability models that are
currently being developed, focusing on describing the development of a
library of a particular class of graphical models called plot models. We
also briefly describe one of the entries in such a library - a model of a
terrorist vehicle attack.

In Section 4, we proceed to detail a protocol we have developed through our
experiences developing this kind of library. This preserves the security of
the data, missing to the external guiding team of academics, whilst still
providing a conduit for fast technological transfer from the academic team
to police. In Section 5, we discuss, in more technical detail, why this
application of graphical Bayesian modelling is feasible and the
circumstances under which formal inferences across the whole library can be applied
to enhance the propriety of this technological transfer. In particular, we
outline how partial information retrieved from collections of past
incidents can sometimes be covertly applied by police to a current incident in an
entirely formal and justifiable way. The BN framework thus enables academics
to guide police in their accommodation of the types of systematic
missingness of informedly censored data that often dominates the open source
information space. This also presents a systematic methodology for building and relying upon libraries of models for different incidents of recurring types of plot to overcome the challenges faced as a result of data about ongoing plots not being readily available to support real-time decision making. We end the paper with a short discussion of ongoing work
and future challenges.

\section{Bayesian Analyses for Secure Domains}

\subsection{A Probabilistic Foundation for Framing Protocols}

Historically, a common way for policing bodies to commission help in building
probability models of sensitive domains has been to simply guess what an
incident in a more benign domain analogous to a sensitive modelling problem
of interest might be. Then, by copying the methodology which applies to this
benign domain, police build their own in-house model of their more sensitive
applications. Of course, matching across domains in this way is perilous for
any party who only partially understand the academic domain they match to.
There may exist no such match. And even if a good match to the secure domain
is successfully identified, the subsequent transfer of technology is
often in practice na\"ive because that transfer has needed to be exploited
entirely in-house and is open to little quality control. This has limited
the success of the technology transfer of frontier inferential methodology
onto critical, sensitive policing domains.

However, more recently within the UK and elsewhere, a more direct transfer
of cutting-edge statistical and machine learning technologies, modelling
certain sensitive applications, has been commissioned. This has involved a
sustained nurturing of suitable academic relationships across a
firewall. A technical team within the relevant policing organisation work
symbiotically and in real time alongside an academic team. The academics
learn as much as they can about the domain through open source
information and create a model from this. They then translate this particular model and
associated inferential methodologies - \emph{bespoke to this particular
domain} - to help police build a customised probability model suitable to
inform the decision support tools they need to protect the public from harm.

The fact that the academic team understand the actual domain of interest as far
as security allows means that they can not only help police template
their models so that they are fit for purpose, but also - through
understanding the types of algorithms that police are using - that they can provide vital
ongoing support to ensure an appropriate calibration and adaptation of these
models as the crime environment develops. We note that a significant recent
investment of resources within UK policing organisations has now
ensured there are sufficient in-house skills available to make such
co-creation possible.

Although such co-creation projects for fast technological transfer continue to
improve, through active engagement over a number of years and over a
diverse set of domains, we have been able to begin to develop protocols
through which this symbiosis might work. It is therefore timely to share
these. This paper describes one such development - the building of
graph-based decision support systems to pursue criminal plots.

The particular approach we describe here is a subjective Bayesian one. There
are a number of reasons for this choice:

\begin{enumerate}
\item Bayesian decision support systems are \emph{prescriptive} in nature.
These therefore almost automatically carry with them a systematic framework
around which a protocol for technological transfer we have outlined above
can be performed.

\item Bayesian methodologies are now \emph{widely developed} and arguably
provide the best modelling framework for a spectrum of different challenging
inferential settings. More specifically, these have now been successfully
applied across a myriad of complex domains very similar to various secure
domains on to which technologies\ need to be transferred.

\item Bayesian methods \emph{interface well with cutting-edge data analytic
methods} - currently being developed in both computational statistics and
machine learning communities.

\item A critical feature of many secure environments is that the
streaming time series and historic data often suffers from being systematically
missing not at random, disguised and, for some central variables, completely
latent. This is typically the case for high-threat criminal plots for which data about the plot is usually only observed and becomes available after the plot is carried out or prevented. This usually demands that expert judgements need to be explicitly
embedded into models before these are fit for purpose. The Bayesian paradigm
provides \emph{a formally justifiable way for embedding such necessary
expert judgements} into that inferential framework through the use of priors that can be updated with further judgements and data as they become available.

\item Once such prior probabilities are elicited, Bayes Rule and  graphical propagation
algorithms can be embedded in the code to update police
beliefs about the underlying processes even when that data is seriously
contaminated or disguised - in ways we illustrate in this paper.
In this way, Bayesian methodology transparently informs and helps police \emph{adjust their
current beliefs} in terms of what they observe which, as well as being consistent
with the way they make inferences, \emph{puts the police centre
stage}. The approach therefore helps them \emph{own} the support given by
the academic team.
\end{enumerate}

Recall that to perform a Bayesian decision analysis (see e.g. \cite{smithbook}), for each \emph{decision} $d\in \mathbb{D}$ - where henceforth the decision space
$\mathbb{D}$ includes the option $d_{\phi}$ of doing nothing - the
policing organisation calculates their subjective expected utility (SEU) $
\overline{U}(d)$ to identify the option $d^{\ast }\in \mathbb{D}$ that
maximises this function. When the police team have a utility function $U(%
\boldsymbol{a})$ on a vector of attributes $\boldsymbol{a}$, and $p_{d}(%
\boldsymbol{a})$ denotes their probability density over their attributes $%
\boldsymbol{a}$, these SEU scores are given by

\begin{equation}
\overline{U}(d)\triangleq \int U(\boldsymbol{a})p_{d}(\boldsymbol{a})d%
\boldsymbol{a}  \label{expected utility}
\end{equation}

In a criminal setting, the attributes $\boldsymbol{a}$ will typically include
measures of harm to the public as well as measures of the policing resources required to
counter this threat in various ways. Within this paper, we will assume, for
brevity, that these attributes have been elicited and known to all parties.
Discussions of how each attribute could be elicited have now appeared
elsewhere \cite{Shenvi2023, Ramiah2024}. Under
this assumption, police will then simply need guidance to appropriately
construct the joint probability mass functions $\left\{ p_{d}(\boldsymbol{a}
):d\in \mathbb{D}\right\} $ - the subject of the remainder of this paper -
so that they can calculate the scores $\left\{ \overline{U}(d):d\in \mathbb{D}\right\}$ (from equation \ref{expected utility}) needed to score the efficacy of any potential
interventions they might make to mitigate harm to the public.

Usually, police will need to learn about $\boldsymbol{a}$ indirectly through
gathering information about a much longer random vector $\boldsymbol{\xi =}%
\left( \xi _{1},\xi _{2},\ldots ,\xi _{n}\right) \boldsymbol{\ }$with
probability densities mass functions $\left\{ p_{d}(\boldsymbol{\xi })\in 
\mathbb{P}_{\mathcal{G}}:d\in \mathbb{D}\right\} $. 
The
components of $\boldsymbol{\xi }$ measure the critical explanatory features
of the underlying process of the criminal activity as understood by the
police. Although our protocols apply more widely, for simplicity - and for
consistency with the topic of this special issue - we will assume in the
first instance that the relationships between the nodes of the vector $%
\boldsymbol{\xi }$ can be described by a two time slice dynamic Bayesian
network (2TDBN) \cite{KorbNicholson2011}. Note that it is trivial to check that any 2TDBN with a finite number of time steps can be written as a
probabilistically equivalent BN through displaying separate nodes for each variable at each time step, and drawing edges for time-lag relationships, as well as for those existing in the static structure.

\subsection{Eliciting a Bayesian Network for Sensitive Domains}

\subsubsection{Four elicitation steps and some causal hypotheses.}

When the vector $\boldsymbol{\xi }$ is not short, applied Bayesian modellers
have discovered that it is wise to build $\left\{ p_{d}(\boldsymbol{\xi }%
):d\in \mathbb{D}\right\} $ in four steps \cite{Wilkerson2021}.
The first step is to elicit those variables that best describe the criminal
process in focus. This step is often missing in standard BN analyses of data
rich environments where interest lies solely in making inferences about the
relationships between various \emph{prespecified} measurements. However, when
describing a type of crime, this step is critical; much of the expert
judgements from police about how crimes unfold is embedded in the choice of the
components $\boldsymbol{\xi }$ and forms a central theme within the
co-creating protocol we describe below. These critical features $\boldsymbol{%
\xi }$ - usually latent to the police team - will be used to frame
hypotheses about what might actually be happening within an unfolding
current suspected incident. But police will often only be able to
observe the out-turns of these features, only knowing a task is being performed once it is complete. Available data does not usually directly inform
these features, and, when it does, it is often ephemeral and disguised.
Furthermore, even what can be observed, often by its very nature, cannot be
shared with the academic co-creation team without betraying secret police
capabilities. On the other hand, interpretable and explainable latent variables -
becoming some of the components of $\boldsymbol{\xi }$ - can usually be
freely shared because these and their relationships are generic, and data and
expert judgements informing these are typically available in the open source social science domain.

Secondly - once the components of $\boldsymbol{\xi }$ have been
identified - the academic team can begin to elicit structural prior
information from police about the likely types of relationships between
them, here assumed expressible as a BN $\mathcal{G}$ whose vertices are the
components of $\boldsymbol{\xi .}$ Again, because of the paucity of data on
many complete past incidents across the whole system, a class of possible
BNs is usually elicited \emph{before} accessing past data sets. Then, using
standard Bayesian technologies, any informative data streams can be
subsequently used to adjust such prior structural hypotheses. This
structural elicitation phase then forms the first stage of the domain
elicitation. The graphical model structure ensures that domain expertise,
predominantly expressed through natural language by police, can be embedded
at the very core of the stochastic model by the academic team. This critical
faithful qualitative information is supported by reasoned arguments, and so
is typically much less ephemeral than its probabilistic embellishments and
is securely placed at the heart of the police team's predictive models. This is
especially important both in classifying different categories of incident
and for describing the types of driving processes behind individual
incidents within each category.

The third stage of the elicitation process is then for the academic team to
guide police in embellishing this qualitative model into a set of full
probability models needed for a Bayesian decision analysis. This is done by
first eliciting prior information - here about the probability that a
particular BN best describes a particular class of plot and then the
particular CPTs of each given category of incident, given each type of
suspect and environment, and each possible intervention police might make. These
expert judgements are then calibrated against any available historic data
relevant to the category of crime being set up - often sadly sparse for many
of its CPTs. For example, CPTs capturing the intent and
capability of a suspect are usually a central part of a BN of an unfolding
crime. However, a criminal's intent will only be fully known by the perpetrator
and data about the capabilities of any particular suspect is likely to be
missing. This is why probabilistic expert judgements from criminologists and
police, and appropriate statistical models, need to inform such tables. Of
course there will be considerable data collected on each past case that
inform some of the entries in the CPTs of the relevant BNs. So once these
expert judgements are in place, such probability judgements can be further
refined using the usual Bayesian machinery designed to do this.

Thus only once this probability model is in place are the final conditional
probabilities elicited that are needed to complete the CPTs
for the current incident involving a particular triaged suspect.
Then, using this probability model as a prior for the unfolding incident,
any available streaming data collected that informs the current incident can
be used to update the predictives about the progress and potential outcome
of the incident currently being policed. This is achieved simply through
police using customised propagation algorithms matched to those provided by
software transferred by the academic teams using their parallel, less
informed BN and its CPTs.

\subsubsection{Causality and libraries of crime models}

One critical issue is, as far as is possible, for the academic team to try
to ensure that the elicited BNs of each category of criminal process are 
\emph{causal}. Fortunately, BN representations of causal processes have been
widely studied for a number of decades, both from a foundational and
methodological perspective beginning with seminal work by \cite{Pearl2000,Spirtes1993}. The majority of this development has focused on the
development of causal discovery algorithms. However, the established
reasoning machinery it utilises can also be applied to build Bayesian models
of the type we need here \cite{KorbNicholson2011, smithbook}.

To apply causal reasoning to criminal processes, in \cite{Ramiah2024}
we argue that it is helpful to embed further properties of a model which
might justify it being described as truly causal. These properties were
demanded long ago by \cite{Hill1965} but have largely been ignored
until recently by the graphical machine learning community. The three
causal properties we demand in the construction of a BN that models a
particular category of crime we discuss in this paper are given below:

\begin{enumerate}
\item The BN provides a framework not only for police predictions about what
will happen if they do not intervene, but also if they do \cite{Pearl2000,Spirtes1993,smithbook} - a property we call \emph{interventional causality} - customised so that these apply in an adversarial
setting \cite{Ramiah2024}.

\item The chosen BN provides a template for the way many different crimes
within a given category might unfold. So this must have this type of generic
quality called \emph{causal consistency} in \cite{Hill1965}. We note
that this type of concept has recently reappeared in a rather different form
as abstraction transport \cite{Felekis2024}.

\item To double guess a criminal's reactions to an intervention they might
make, it would be helpful if the police team tried to ensure that the structural
beliefs expressed through the graph were shared by the criminal \cite{SmithAllard1996, Ramiah2024}. This is a strengthened version of the long-established \textit{coherence} property that we refer to as \emph{causal
common knowledge.}
\end{enumerate}

In a causal model as described by the first bullet above, the structural
framework of its BN will be shared for all $\left\{ p_{d}(\boldsymbol{\xi }%
):d\in \mathbb{D}\right\} $ and, by the second bullet, for any given incident
within the category. We note that the first invariance property is now widely
hypothesised for the BN/2TDBN in order to make inferences about, for
example, the efficacy of treatment regimes, albeit in a non-adversarial
setting - see e.g. \cite{KorbNicholson2011}.

The most studied class of interventions $d_{I(\widehat{\boldsymbol{\xi }})}\in \mathbb{D}$ associated with an interventionally causal BN are ones
which force the measurement $\xi _{i}$ to take the value $\widehat{\xi }_{i}$
for each $i\in I\subseteq \left\{ 1,2,\ldots ,n\right\} $ with $\widehat{%
\boldsymbol{\xi }}\triangleq \left\{ \widehat{\xi }_{i}:i\in I\right\} $ -
interventions commonly referred to as "doing" $\left\{ \xi _{i}=\widehat{\xi 
}_{i}:i\in I\right\} $ (see \cite{Pearl2000, Spirtes1993, smithbook}). Then the valid BN for the intervention is one which simply substitutes the CPT of
the conditional mass function of $\xi _{i}$ with one that assigns
probability one to the event $\left\{ \xi _{i}=\widehat{\xi }_{i}\right\} $
irrespective of the parent configuration whilst leaving all other CPTs the
same as they were when there was no intervention. Pearl calls a BN $\mathcal{G}$ \emph{causal} when these BNs, perhaps embellished with different collections of CPTs within the same directed acyclic graph (DAG) $\mathcal{G}$, $d_{I(\widehat{\boldsymbol{\xi }})}\in \mathbb{D}$,
are all valid assertions about a given domain \cite{Pearl2000}. Then the joint density $p_{d\mathcal{G}}(\boldsymbol{\xi })$ for all $d_{I(\widehat{\boldsymbol{\xi }})}\in \mathbb{D}$ can be factorised as:

\begin{equation}
p_{d\mathcal{G}}(\boldsymbol{\xi })=\prod\limits_{i=1}^{n}p_{id}(\xi _{i}|%
\boldsymbol{\xi }_{Q(i)})  \label{causalBN formula}
\end{equation}%
where $\boldsymbol{\xi }_{Q(i)}$ is the set of parents of $\xi _{i}$, $%
i=1,2,\ldots ,n$ in $\mathcal{G}$. Within the adversarial setting of
criminality, such vanilla assumptions will not usually hold for a simple BN
of the unintervened process \cite{Ramiah2024}. However, we demonstrate
in \cite{Ramiah2024} that - provided the BN is chosen to be rich enough to model the capability of the criminal and their intent, and what
the criminal might be able to learn about how the intervention might be
made - then the same DAG structure and analogous
substitution rules can be used. Police will then simply
substitute some of the CPTs valid when no intervention is made for others
when implementing $d\neq d_{\phi }$, $d \in \mathbb{D}$ where the particular
substituted CPTs made are determined by $d$ can
be assumed valid in this adversarial context. Examples of the precise nature
of such substitutions is beyond the scope of this paper but can be found in 
\cite{Ramiah2024} and \cite{Ramiahplot24}.

This interventional causal property for a BN model of a policing application
can be particularly useful because the same BN can then be used as a
predictive framework whether or not they choose to intervene to try to
prevent a crime being committed. We note that the BN elicitation protocols
developed by \cite{KorbNicholson2011} try to ensure that this type of causal property is
automatically embedded within the model. For the purposes of this paper, we
henceforth assume that this type of causal property is co-created by the
two teams for all interventions $d\in \mathbb{D}$ that police might consider making
for a criminal process described by its graph $\mathcal{G}$.

For the purposes of building a \emph{library of BNs} for particular
categories of crime that can then be used to match similar yet distinct incidents,
we also need to demand both the \emph{consistency} and \emph{coherence} properties that were first demanded
of causal systems by \cite{Hill1965} and are here applied to BNs. So we
require that the elicited BN will \emph{remain a valid template for many
analogous instances} of crime within the same category \cite{Wilkerson2021,Ramiah2024}. A necessary skill of academic teams
is the ability to ensure that categories are defined sufficiently finely
so that they are similar in this structural sense such that the protocol below
applies.

The final property we also may need to use when applying these models to
predict the consequences of a police intervention is one based on the game-theoretic notion of common knowledge. Bradford Hill  \cite{Hill1965}
demanded that the hypotheses embodied by a causal model should, on
reflection, appear at least plausible to other intelligent people. In \cite{Ramiah2024} we strengthen this hypothesis and then apply it to a
suspect assumed by police to be intelligent. So, in particular, inferences police make hypothesise that the suspected criminal is\emph{\
intelligent and shares the same understanding as the police team about how their
planned crime might be successfully perpetrated}. We note that this
hypothesis is only needed if the suspect can learn that police have committed
to intervene in a particular way and can react to this intervention. This is indeed often a true characteristic of criminal activities in practice. Examples of such
visible interventions are police raids or the establishment of protections
for potential targets. The plausibility of this assumption needs to be
tested on a case-by-case basis; in the running example we give below, it is almost automatic. We demonstrate in \cite{Ramiah2024, Ramiahplot24} through a number of examples how
this hypothesis facilitates police, guiding them in double guessing how a
criminal might react after hearing that a particular intervention has been
put in place, and therefore guiding them in producing the necessary forecast distributions associated
with applying interventions that become visible to a criminal.

Most sensitive policing domains are ones where a crime develops dynamically.
However, the BN framework extends into a model of evolving domains where,
in particular, analogous simple "do" algebras are especially simple to
define for the 2TDBN $\mathcal{G}$ over time steps $t=0,1,\ldots ,T$ we use
in our running examples. This is because any 2TDBN $\mathcal{G}$ is
equivalent to a BN with graph $\mathcal{G}^{+}$ whose vertices are time
indexed. Its irrelevance statements are therefore implicit in the DBN - see
e.g. \cite{KorbNicholson2011} for a precise definition of this construction. We can now
duplicate the algebra defined above on $\mathcal{G}^{+}$ and translate this
to $\mathcal{G}$. So such maps fall within our generic framework - albeit
with often a massive set of factors and types of interventions - for example
when an intervention might be applied and for how long, as illustrated in 
\cite{Ramiah2024} and \cite{Ramiahplot24}.

\subsection{Data and Information in Secure Domains}

Under recent co-creation schemes with policing agencies, it has been possible
for academics to help in-house domain experts to better design probabilistic
models of serious crime. However, of course, these academics still only have
restricted information about the domains of application. Several of the
model's attributes will have no data available, either from historic analogues or within the
currently monitored case, to directly inform their CPTs, and a lot of the data that \emph{is} available to police must necessarily be kept secure behind a firewall. Typically, academic collaborators only have available to them:

\begin{itemize}
\item \emph{Criminologists' and sociologists' models of criminal behaviour}
that lie within the open domain. These are especially important because they
often give a great source through which to both categorise different classes
of crime and describe their development.

\item \emph{Open source data about analogous past incidents}. These
typically appear in articles by journalists and within scholarly case
studies of specific events written by criminologists. Although this is often
not data in a statistical sense, each such report can give information about
the development of past instances within a particular category and so inform
the CPT of an associated BN.

\item \emph{Access to someone from behind the firewall}. Such a person will
be free to disclose relevant, less sensitive domain information that might
begin to fill out newly arriving information necessary to build both the
structure and the probability factors of a probability model with sufficient
specificity to be part of a Bayesian decision support system.

\item \emph{Securely emulated data} generated though in-house software unknown to
the academic co-creator, calibrated with secure inputs. This has proved to be a valuable tool
for checking predictive algorithms provided by academic teams - although of
course such tests can only be as good as the outputs of the emulation tool
used to test it.
\end{itemize}

It has been established through work in less secure domains that - once the
challenges associated with sparsity of data on at least some variables has
been acknowledged and addressed - various graphical models provide the ideal
framework around which to build well-calibrated models behind firewalls. The
sources described by the first and third bullet inform the early structural
phases of the modelling processes, whereas the second and third bullet inform the
embellishment of the model. Furthermore, the last three bullets enable the
model to be tested and refined into a working piece of code which remains
securely embedded behind a firewall.
\section{Criminal Plots}

\subsection{Introduction}

Criminal activities can take on a number of guises. Many crimes are simply
opportunistic in nature and so are less predictable except in a population
sense. But many other crimes - especially serious crimes - need a degree of
planning or preparation. This makes it at least feasible for police to
frustrate the perpetration of individual incidents by appropriately
intervening in their preparation. We have argued in \cite{Ramiah2024}
that different genres of criminal activities each demand their own type of
model. However, we here discuss one particularly interesting broad category of criminal
activity that we have called a \emph{plot} \cite{BunninSmith2021,Ramiahplot24}. Plots can be described by a subclass of the 2TDBN. So a protocol
for establishing a library of plots provides an example of establishing a
library of BNs in a context where data is typically missing not at random
and is often disguised. Here, because the discussion of terrorist plots is now
open source \cite{BunninSmith2021,Shenvi2023}, our running
examples will focus on the co-creation of this class.

\subsection{Plots as a Hierarchical Bayesian Model}

Plot models - like the ones first discussed in \cite{BunninSmith2021,Ramiahplot24} - have been directly elicited from domain experts and
can be expressed as a 2TDBN. This BN graphically expresses a 3-level
hierarchical model of a plot's description whose lowest two levels are
usually latent:

\begin{enumerate}
\item At the deepest level of this hierarchy is a latent discrete Markov
process modelled by a time series of random variables $\mathbf{W}%
^{T}\triangleq \left\{ W_{t}:t=1,2,....T\right\} $. This discrete time
series indicates in which of a number of preparatory phases - elicited from
domain experts to characterise a particular class of crime - the given plot
might lie at any given time. Represent the phase of a particular
plot at time $t$ by the indicator variables $\left\{ w_{0t},w_{1t},w_{2t},\ldots ,w_{mt}\right\} $, $t=1,2,....T$. The particular phase denoted by $w_{0t}$ is called the \emph{%
inactive} phase - an absorbing state where the criminal has aborted their
plot. The other phases we call \emph{active} phases. The time series of an
incident's phase is typically latent to police - although insider
information or occasional revelations might directly inform it.

\item Within a plot, when in a particular phase, a criminal will need to
complete certain tasks - characteristic of a certain class of crime - before
they can transition to a subsequent phase. At each time step, this intermediate
layer of the hierarchical model is a \emph{task vector} $\boldsymbol{\theta 
}_{t}=\left\{ \theta _{1t},\theta _{2t},\ldots ,\theta _{nt}\right\} $
consisting of component tasks. These are often indicator variables on
whether the criminal is engaged in the given task or not. Subvectors $
\boldsymbol{\theta }_{tI(w_{j})}$, called \textit{task sets}, of the task vector $\boldsymbol{\theta }%
_{t}$ are defined as those tasks whose marginal distributions for a given active phase $w_j$, $%
j=1,2,\ldots ,m$, are distinct from their distributions for the inactive phase $w_{0}$. So these activities suggest
that the suspect might be in the phase $w_{j}$. Let $\boldsymbol{\theta }%
^{T}\triangleq \left\{ \boldsymbol{\theta }_{t}:t=1,2,\ldots ,T\right\} $.
Again, the components of $\boldsymbol{\theta }^{T}$ will typically be latent
and only inferred by police, although, on occasion, police might happen to
directly observe that a particular task is underway or complete.

\item Police will typically have routinely available to them streaming time
series of observations called \emph{intensities} $\boldsymbol{Z}%
^{T}\triangleq \left\{ \boldsymbol{Z}_{t}:t=1,2,\ldots ,T\right\} $ about
the progress of a suspected plot. The components $\boldsymbol{Z}_t=\left\{\boldsymbol{Z}_{1t},\boldsymbol{Z}_{2t},\ldots,\boldsymbol{Z}_{nt}\right\}$ of
the intensity vector are chosen to help police discriminate whether or not
an agent is engaging in the task $\theta _{it}$, $i=1,2,\ldots ,n$ at
time $t$, $t=1,2,....T$. We illustrate this below. These are, by definition,
seen by police. However, for any ongoing incident, academics outside the
firewall will typically not have access to the values of these data streams,
at least not until the criminal has been convicted. So often the components
of $\boldsymbol{Z}^{T}$ and nearly always the values they might take within
an ongoing incident will be missing to the academics guiding police support.
This information is highly sensitive because criminals could disguise the signals
they emitted or even distort these to deceive their observers if they learned what
police could see of their activities.
\end{enumerate}

For $t=1,2,\ldots ,T$, the 2TDBN\ will have as its vertices the components
of $\left( \boldsymbol{\xi }_{t-1},%
\boldsymbol{\xi }_{t}\right)\triangleq(\boldsymbol{W}_{t-1}^{t},\boldsymbol{\theta}_{t-1}^{t},\boldsymbol{Z}_{t-1}^{t})$ where: 
\begin{equation*}
\boldsymbol{W}_{t-1}^{t}\triangleq \left\{ W_{t-1},W_{t}\right\},
\boldsymbol{\theta }_{t-1}^{t}\triangleq \left\{ \boldsymbol{\theta }_{t-1}, \boldsymbol{\theta }_{t}\right\} , \; \text{and} \; \boldsymbol{Z}_{t-1}^{t}\triangleq\left\{\boldsymbol{Z}_{t-1},\boldsymbol{Z
}_{t}\right\}.
\end{equation*}
Full details of the factorisations of the joint density of $\boldsymbol{\xi}$ are given
in \cite{Ramiahplot24} which is consistent with a standard 2TDBN with
Markov time slice graph $\mathcal{G}_{t}$, $t=1,2,\ldots ,T$. The graph of one such 2TDBN is given below. The full BN $\mathcal{G}^{+}$ then simply concatenates these graphs together. Note that, by definition, all vertices
of a 2TDBN which are components of $\boldsymbol{\xi }_{t-1}$ are founder
vertices - i.e. have no parents.

For plot models, the vertex $W_{t}$ has the single parent vertex $W_{t-1}.$
The subgraph $\mathcal{G}_{t\theta }$ of $\mathcal{G}_{t}$ generated by the
components of $\left( \boldsymbol{\theta }_{t-1},W_{t},\boldsymbol{\theta }%
_{t}\right) $ is an elicited DAG on the vertices drawn
from components of $\left( \boldsymbol{\theta }_{t-1},W_{t},\boldsymbol{%
\theta }_{t}\right) $ describing the generating latent process. The
components of $\left( \boldsymbol{\theta }_{t-1},W_{t}\right) $ in $\mathcal{%
G}_{t\theta }$ are founder vertices. The parents of the components $\theta
_{it}$ of $\boldsymbol{\theta }_{t}$ must include $W_{t}$ and all components 
$\theta _{i^{\prime }t}$ $i^{\prime }<i$ where both $\theta _{i^{\prime }t}$
and $\theta _{it},$ for some phase $w_{j}$, are components of the task
vector $\boldsymbol{\theta }_{tI(w_{j})}$, $j=1,2,\ldots ,n$, but otherwise any DAG may be valid. 

\begin{figure}
    \centering
    \includegraphics[width=0.63\linewidth]{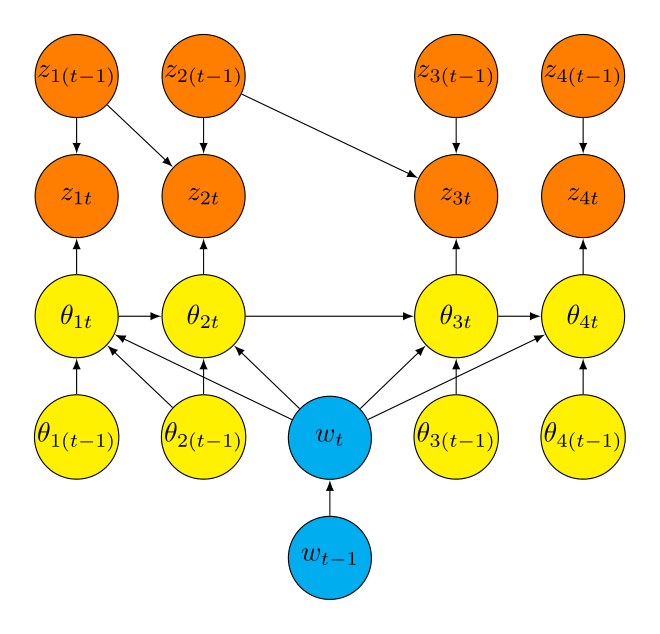}
    \caption{Example 2TDBN for a plot model with four tasks with phases coloured blue, tasks coloured yellow and intensities coloured orange.}
    \label{fig:2TDBN}
\end{figure}

Finally, the subgraph $\mathcal{G}_{tZ}$ of $\mathcal{G}_{t}$ generated by $%
\left( \mathbf{Z}_{t-1},\boldsymbol{\theta }_{t},\boldsymbol{Z}_{t}\right) $
has as founder vertices components $\left( \mathbf{Z}_{t-1},\boldsymbol{\theta }_{t}\right)$. For a plot model, we simply demand that, for each $i=1,2,\ldots ,m$, the component $%
Z_{it}$ must have as a parent only $\theta _{it}$ out of the task vector, but it can have an edge from any of the components of $%
\mathbf{Z}_{t-1}$ indexed before it. This is ensured by defining $\theta
_{it}$ so that it directly informs the task $\theta _{it}$ alone, $%
i=1,2,\ldots ,m$. There are no edges from $W_{t-1}$ or $W_{t}$ to any
component of $\boldsymbol{Z}_{t}$. This is because the
intensities and tasks for plot models are defined in such a way that observations can inform
the phase of a plot only through the tasks the criminal engages in to pass
through that phase. An example of a graph $\mathcal{G}_{t}$ of an 2TDBN valid for all times $%
t=1,2,\ldots ,T$ on the components of $\left( \boldsymbol{\xi }_{t-1},%
\boldsymbol{\xi }_{t}\right) $ of a simple elicited plot when there are only four tasks is given in Figure \ref{fig:2TDBN}.

There are several additional structural properties demanded of a plot model
that are not embedded in a 2TDBN. These involve hypotheses about the formation of the task set, the
impossibility of passing from some phases to others and the fact that many
entries in different CPTs within the subgraph $\mathcal{G}_{t\theta }$ must be identical. This
encouraged us in \cite{Ramiahplot24} to develop a bespoke class of
graphical model that could graphically express all the structural hypotheses
associated with this class. However, the more familiar 2TDBN is still a valid
framework for inference and therefore ideal to illustrate the secure
elicitation protocol we outline below as this might apply to a BN.

\subsection{Causal Modelling for Decision Support on Plots}

For a plot model, because its variables explicitly describe how and
why the crime develops the way it does, it is almost immediate that it is
interventionally causal. It describes the precise processes under which an
attack might unfold - not just the likely way variables might depend on each
other. It therefore transpires that it provides an excellent template for a
causal 2TDBN \cite{Pearl2000}. A detailed discussion of this
point is given in \cite{Ramiahplot24}. We also demonstrate that
it is consistently causal in the sense that the topology of the 2TDBN of one
category will apply to most instances within it, as will many of its CPTs.
This is another reason for using a library of plot models as our running
example.

However, when the criminal can become aware of at least one of the
interventions police might consider adopting, typically the topology of the 2TDBN needs to be more expressive than if this were not the case. If the suspect cannot foresee an intervention from police, they will never deviate from their \textit{modus operandi} (MO), hence no alternatives need to be modelled. If they \emph{can} foresee interventions, they may deviate from their MO and police must consider these deviations in their model. Two different detailed illustrations of how these embellishments can be systematically embedded in a framing graph are given in \cite{Ramiah2024} and \cite{Ramiahplot24}.


\subsection{Graphs in a Library of Terrorist Plots}

\subsubsection{An example of a library of plot models}

Suppose we are developing a library of graphs that might support the
real-time decision making of police charged with defending the public from
various forms of terrorist outrages from a lone attacker. For the sake of
simplicity we will assume that, at least in the first instance, the police team
want to incrementally co-create a decision support tool that can help them
defend against one of five basic forms of lone attack - a knife or sword
attack, a crossbow attack, a firearms attack, a vehicle attack and a bomb attack. We will assume that we are at the initial point of co-creation where
we plan the first entry within the library - a vehicle attack by a category of
known suspects - here, a planned vehicle attack orchestrated by IS. We
provide a little detail of this context below.

Police and criminals are both aware that
terrorist plots of the five types
mentioned above typically progress through a set of phases, all of which
need to be completed before any such attack is successful. The generic
phases of this progress can be listed as follows: $w_{0}$ - not engaged; $%
w_{1}$ - recruited to the plot; $w_{2}$ - training to be capable of
perpetrating the plot; $w_{3}$ - identifying an appropriate target of the
plot and reconnoitring it; $w_{4}$ - obtaining the equipment needed to
attack; and $w_{5}$ - travelling to the target to make the attack. It is good
practice when eliciting a probability model that might translate into
other applications to define states as generically as possible in order
for this translation to be made as straightforward as possible. We note that we
have subsequently found that, for the majority of types of attack in
libraries of plots like the ones described above, these phases define the
states of the phase variables $\left\{ W_{t}:t=1,2,\ldots ,T\right\} $
sufficiently finely to provide the predictive capabilities police need. So
in fact both for a vehicle attack and other attacks above, $\left\{
W_{t}:t=1,2,\ldots ,T\right\} $ can be assumed to take one of these six states.

On the basis of open source data, academic teams have learned that the
transition matrices between these six states will, however, be a function of
which of the five types of attack the orchestrator chooses, the intent of the
perpetrator, their capabilities, and the quality of the defences police
might employ. For some classes of plot, such as exfiltration plots, some of
these phases could occur simultaneously. In this case we have needed to
refine the states in an obvious way so that these are disjoint. However, in
terrorist plots, given the categories of type, intent, capability and defence
describing a particular suspect than their environment, and especially with
the perpetrator acting alone, it is usually safe to assume that the
suspect lies in just one of these states at any given time. Henceforth for
the sake of simplicity assume this is the case for all examples we give here.

Here already we can make some plausible assumptions about any one of the five
types of plot. For example:

\begin{enumerate}
\item At any time, a potential perpetrator may choose or may be forced to abort a
plot and so transitions into the absorbing state $w_{0}$. This means that there
is a benefit to police making it more difficult for the terrorist to
transition through the required phases to be in a position to perpetrate a
crime.

\item Clearly by definition a suspect must have been recruited in the past -
state $w_{1}$ - before transitioning to later states.

\item Once a given suspect is skilled up to be able to perpetrate a plot -
i.e. passing through $w_{2}$ - within the time frames police would be
working in, that suspect will remain trained and will not lose these skills.

\item However, the suspect can review and substitute one identified target ($%
w_{3}$) with another or can equip themselves or discard the equipment ($%
w_{4}$) at any
time they choose. So just because a suspect has been in the phases $w_{3}$
or $w_{4}$ at one time does not mean they are currently equipped or have a current target identified.

\item The perpetrator cannot attack - phase $w_{5}$ - until all other
active phases have been completed.
\end{enumerate}

So once the nature of a particular attack has been elicited, many of these
structural elements of one model within the library - here concerning how a
suspect might transition the phases of the plot - can be immediately shared
and translated to other entries in the library. Note that it is not possible
to embed these structural hypotheses explicitly within the topology of the graph of a
2TDBN, so these will need to be logged by academics and embedded implicitly
within this structure. These structural assumptions can nevertheless greatly
simplify the quantitative elicitation process.

We often find that some idiosyncratic structural features will often need to
be embedded too. For example, the order a suspect chooses to pass through the
phases $w_{3}$ and $w_{4}$ will depend on how difficult the equipping phase $%
w_{4}$ might be \cite{SmithShenvi18}. This will typically depend on the
nature of the plot as well as the capabilities of the suspect.

In contrast to the largely shared structural expert judgements of police,
the collections of tasks associated with different types of terrorist
attacks will usually differ significantly across different entries. For
example, the tasks undertaken to ensure that a suspect is sufficiently skilled and trained to
perpetrate a vehicle, bomb, firearm or knife attack are quite different
from each other - as are the tasks needed to equip themselves for such
attacks. Nevertheless, some tasks will often be shared across varying types
of plot - especially those associated with communications across
co-conspirators and the identification of appropriate targets. This means
that some of the entries of CPTs associated with different entries within a
library can be shared.

When it comes to intensities, the precise nature of what police might have
available to them will often not be known to the academic team. It is in
this aspect that the academic stochastic model and the one police use
behind their firewall could be very different from one another. We will see,
however, that, because of the underlying structure of the plot model, the
actual updating algorithms for both models will be identical or very similar.

Furthermore, there are very obvious types of observation that academics can
speculate will be available to police and then build their academic model in light of these speculations. For example, web searches for hiring a heavy
goods vehicle are clearly indicative that a suspect is trying to arm
themselves for a vehicle attack. So - based on common sense and open source
information about what police \emph{might} have available to them - the
academic team can make informed conjectures at least about the broad nature of
this genre of information, although possibly not its reliability and
certainly not the values such information might take in any live case.

Some such speculation will be accurate, others less so. Nevertheless, police
behind the firewall will be able to weigh the plausibility of any academic
speculation and adapt their internal code to correct within their own
replica system any poor guesses made by academics. The initial attempts by
the academic team therefore guide the police team in modelling the real
process through providing a template for how any data might be embedded in a
model. The original academic model also demonstrates to police how
valuable certain data streams might be.

\subsubsection{An example of a graph in a library - vehicle attacks}

The following model of a vehicle attack plot made by a lone IS
attacker has already been reported in open literature (see
e.g \cite{BunninSmith2021}) so can be shared here. Because for this early
co-creation the protocol described below was only nascent, describing how we
would now proceed becomes hypothetical and will not betray any actual
potentially sensitive co-creation between the two teams.

Within this co-creation, the academic team learn that this type of plot
typically progresses through a set of phases outlined above for a triaged
suspect who has become a person of interest. The academic team are then
pointed to open source literature which explains that the typical tasks $%
\boldsymbol{\theta }_{tI(w_{1})}$ a suspect might engage in when in the
first active radicalisation phase described above are typically shared with
the other plots in the library. These include the individual tightening
their relationships with like-minded people and a progressive retreat from
day-to-day contacts with otherwise close contacts such as certain friends or relatives who might strongly disapprove of the plan.
Academics therefore learn from open sources that by monitoring the suspect's
web activities and social media, through phone logs and through learning
about recent interactions with known IS sympathisers using CCTV that the
intensity of such activities will all inform police about the likelihood
that the given suspect is within this phase, or was in this phase at some
time in the past. Therefore, they can conjecture that measurements of these might be
the tell-tale signs (intensities) police might use as subvectors of
components $\boldsymbol{Z}_{I(w_{1})}^{T}$ of $\boldsymbol{Z}^{T}$. Although
police capability to harvest such information within suspected incidents
will be highly confidential, the academic teams can nevertheless create for
themselves synthetic data streams about various hypothetical incidents and
capabilities that can demonstrate how the statistical model might inform the
police about whether and when this phase was enacted. Early illustrations of
this process appear, for example, in \cite{BunninSmith2021} and elsewhere
in other police handbooks.

To realise the plot, a suspect who cannot yet drive a heavy goods vehicle
would then need to somehow increase their capabilities to learn how to do
this - a preparatory step we denote by phase $w_{2}$. Associated tasks in the
middle layer of the hierarchy might be to sign up for a commercial vehicle
training course or to learn this skill from an accomplice already able to drive
and who has access to such a vehicle - who is perhaps abroad and can commit time to this
activity. All such options at time $t$ - denoted by the task subvector $%
\boldsymbol{\theta }_{tI(w_{2})}$- are again common sense and so are also
obviously available to the academic team. Note that if police have
information that the suspect already has this training then of course they
pass instantaneously through this phase. Their code can obviously be
designed to facilitate the accommodation of such very incident specific
information. 

The academic team could also conjecture plausible signature intensity
measures of such activities available to the police team - for example
discovering the suspect has searched for a commercial driving course or
finding evidence the suspect has attended such a training course.
Alternatively, police might observe searches for travel or booking of flights
to a country where they could receive such training, or might observe
association with a local friend known to have these skills and capabilities.
Other information might be more physical - for example, from CCTV images to
observe the suspect at an airport or driving such a vehicle with a friend.

The suspect will then need to \emph{identify a target} and reconnoitre it -
phase $w_{3}$ -\ where he could cause the most drama and spread the
most fear, individually or working with accomplices. The tasks
associated with this phase are typically common to all types of attack and
have appeared in press reports and scholarly articles and so again are largely
known to the academic team. These include investigating the demographics and
defences of different candidate sites which could involve visiting potential
target sites and/or electronically exploring maps of potential target areas.
These two possible tasks are represented in $\boldsymbol{\theta }_{tI(w3)}$.
Again, the academic team might conjecture that such activities might be
captured by police monitoring the suspect's web data, their interception of metadata associated with phone messages, CCTV and direct observation which could
all form arguments in the intensity function $\boldsymbol{Z}_{I(w_{3})}^{T}.$

Once sufficiently trained to perpetrate the plan, the suspect will also need
to \emph{source the heavy goods vehicle} - phase $w_{4}$ - to use in the attack.
This phase will typically involve either hiring such a vehicle, being given
it, or stealing it. Again the academic team could speculate the tell-tale
sign police might be able to observe. Finally they will then need to drive
this vehicle to the target to perpetrate the attack - phase $w_{5}$. This
action could be observed in obvious ways from, for example, traffic
cameras, direct pursuit and intercepted phone messages. Academics can then
build models that harness and then customise state-of-the-art feature
extraction algorithms to construct putative inferential methods that can be
demonstrated on synthetic data and real open source data from more benign
applications. Police can then use these analyses to template their own
algorithms for use within their own secured system.

The point of discussing the above is to demonstrate how, in a given
instance, the academic team can build up a structural model that is close to
a faithful representation of reality. They can then code up a statistical
model of this type of plot that is very similar to one the police team might like
to implement. What remains conjectural is structural information about
the nature of the intensities possibly available to police and
information they might have about that particular person - for instance, whether or
not they were trained to drive a heavy goods vehicle -  to
model a particular suspect within a particular potential incident. But such information
would anyway need to be customised to the particular incident by police
behind their firewall.

Of course, the relevant CPTs would need to be constructed to demonstrate the
2TDBN. But even then note that many of the quantifications needed for such a
model - for example, how quickly the suspect could reach a target when
driving their acquired vehicle - can be elicited directly by the academic
team based on open source data, or otherwise plausibly guessed.

On the basis of such structural information, academics can build and
demonstrate a plot model of such an attack \cite{BunninSmith2021}. Obviously analogous Bayesian structural
models of other types of terrorist plot attacks listed above could be
constructed by the academic team. Such models can then be donated to police
to help guide the construction of their own more realistic in-house matrices
and algorithms that can be used to support an actual secure decision
analysis to help defend a threatened vehicle attack.

In the next section, we provide, for the first time, a detailed description of how
such a library of 2TDBNs can be securely transferred from an academic team
to an operational police model despite significant amounts of the in-house
data being missing to the guiding academic team.

\section{Co-creating a Library of Plots}

\subsection{Introduction}

Broadly, the constraints for a secure technological transfer of any class of
plots are experienced as follows.

\begin{enumerate}
\item A generic description of plots expressed in the \emph{phase
relationships} in the lowest layer of this hierarchical model lies largely
in the public domain - within sociological \& criminological articles, open
source case studies and information that police experts can freely
communicate. So this is directly accessible to the academic co-creators who
can guide its accommodation into this structure. These helpfully categorise
and explain various plots and the motivation and capabilities of various
different types of potential perpetrators of the plots in focus. Some
generic evidence to inform the generic priors on the CPTs associated with
the transition between stages will also be available. This can later be
refined by police using more secure information available only to
them.

\item Generic information about the \emph{tasks} that need to be completed
to move from one phase to another, and the probabilities linked to
both the choice of the task and their ease of completion associated with
various categories of criminals, is also straightforward for the academics to
elicit from sources mentioned in the bullet above. Again - using
instructions from the academic teams - the CPTs can then be refined by
police using other more sensitive evidence they have about various crimes,
using as priors the probabilities on the CPTs based on open source data.

\item On the other hand, the full extent of the \emph{data streams} police
might currently have available to them indicating which tasks are currently
being engaged in \emph{within any ongoing investigation} can be highly
confidential. For example, if a criminal learned that their messages on the
dark web can be unencrypted, then they can use this to disguise their
messages or deceive police and so be harder to apprehend. However, independent of police and based
on open source data, the academic team can
of course conjecture what these might be as we illustrated above. They can
then communicate this open source model to police, populating the topology
of the BN and the associated CPTs ready for adjustment of the police model
behind the firewall in light of what only they know.

\item The \emph{actual data} that police collect \emph{concerning specific
individuals} is highly sensitive and cannot be shared with the academic
team. Were the current suspect to learn what the police knew about the
progress of any plot truly in progress, they would become far more
dangerous. On the other hand, any personal information about a given
suspect cannot be shared until they have been convicted. If the
suspect is in fact innocent, then, as soon as this has been discovered, any
personal information about them cannot be ethically retained. Such
information might be about the category assigned by police to a suspect, the
nature of the information being collected on them and the values of this
data at any given time. However, there is a rich, although usually still
incomplete, bank of open source information about proven criminals, for
example provided by press releases and court reports. So, based on
academic conjectures and rehashing particular use cases, the academic team can demonstrate how the model
might learn in light of such synthetic data sets and share this with police.
The potential usefulness of the open source code can then be demonstrated as
documented in \cite{BunninSmith2021}.
\end{enumerate}

With these points in mind, we next briefly outline the protocol we have used
in co-creation projects with various policing organisations, not just for
terrorist plots of the kind above, but to build other libraries of plots. We
note that the development of the libraries we discuss below can be
incremental - gradually adding and refining entries in the library to
progressively increase the scope of the decision support tools made
available to police. Even partially populated libraries can be extremely
useful to police, especially if the established entries are models for commonly occurring categories of crime.

\subsection{A Protocol for Co-creating Plot Models}

\subsubsection{Notation and setting} \label{NotationSetting}

Here we describe a generic protocol that reflects our current processes, and that
enables an academic team to co-create libraries of BNs where police are
led to their own Bayesian model to support their SEU decision making when
pursuing a criminal in a way that all information that police need to keep secure
remains undivulged. We first need to set up some notation.

Any particular incident whose progress is being monitored will involve a
certain broad category of suspect (e.g. an IS sympathiser, their age and their history) which
will inform the nature and speed of their progress through the different
phases of a plot through the suspect's intent, potential capabilities and MO. We denote this background information by $\mathcal{S}_{b}$.
The academic team will be able to discover from open source publications how
police may be able to categorise a given suspect, although the nature of such
information may be covert and known only to police. A second classifier
effectively concerns the environment $\mathcal{S}_{e}$ in which the suspect
operates, such as place of residence, which might affect the capability and
ease of aborting a plot - information that is more easily exchanged across
the firewall. Let $\mathcal{S}\triangleq \left( \mathcal{S}_{b},\mathcal{S}%
_{e}\right) $ denote this information. We expect that all CPTs will need to
be indexed by $\mathcal{S}$, although most of these will be shared across
categories $\mathcal{S}$. Because the academic team will only
need to \textit{demonstrate} their code within their own library, they typically need to only code up one such
category. For this reason, we have suppressed this indexing in the
development below. However, in the parallel library developed by police,
they will need to elicit different entries for each CPT for each such
category.

Denote by $\mathbb{L=}\left\{ M_{1},M_{2},\ldots ,M_{k}\right\} $ an
arbitrary library of fully embellished probability models - called \emph{%
entries} - where these entries have been ordered consistently with their
arrival in the library. Here, $M_{i}=\left( \mathcal{G}_{i},\mathcal{C}%
_{i}\right) $ where $\mathcal{G}_{i}$ denotes the DAG of the BN and $\mathcal{C}_{i}=\left\{ C_{vi}:v\in V(\mathcal{G}%
_{i})\right\} $ - with $V(\mathcal{G}_{i})$ the vertex set of $%
\mathcal{G}_{i}$ - the collection of CPTs needed to embellish $\mathcal{G}%
_{i}$ into a full probability model, $i=1,2,\ldots ,k$. In \cite{Ramiah2024} and \cite{Ramiahplot24}, by defining an appropriate causal algebra, we
demonstrate how the academic team can design the graphs $\mathcal{G}_{i}$ so
that these provide a valid framework to describe not only how events might unfold
not only when police do not intervene (decision $d_{\phi }\in \mathbb{D%
}$) but also when they do (decision $d\neq d_{\phi}$, $d\in \mathbb{D}$). So in this sense $\mathcal{G}_{i}$ will provide a
suitable framework to describe the structure of \emph{any }probability model
for all contemplated interventions and category of suspect-environment pair $
\mathcal{S}$.

Despite this useful structural invariance, in order to build a full
probability model for each intervention, police will need to also specify for
each category $\mathcal{S}$ the CPTs 
\begin{equation*}
\mathcal{C}_{di}=\left\{ C_{dvi}:v\in V(\mathcal{G}_{i})\right\} 
\end{equation*}%
associated with such decisions $d\neq d_{\phi }\in \mathbb{D}$ - where we write $C_{vi}=C_{d_{\phi }vi}$. Henceforth we focus on
building the library for a fixed category $\mathcal{S}$. To establish
their library, police will then need to repeat the process below for all
other categories of suspect $\mathcal{S}$.

It is convenient to partition the sets $\left\{ \mathcal{C}_{di}:d\in 
\mathbb{D}\right\} $ of CPTs into the three sets $\{\mathcal{C}_{di}(1), \\
\mathcal{C}_{di}(2), \mathcal{C}_{di}(3):d\in\mathbb{D}\}$, where $%
\mathcal{C}_{di}(1)$ denotes those CPTs whose prior information and
informative data sets can be shared with the academic team; $\mathcal{C}%
_{di}(2)$ those CPTs for which the academic team have some information and perhaps
data informing these but for which police have additional information; and 
$\mathcal{C}_{di}(3)$ those CPTs which the academic team have only scant
information about and which police would plan to overwrite with their own
secret but much more reliable information. Let$\left\{ \mathcal{C}%
_{di}^{\prime }(1),\mathcal{C}_{di}^{\prime }(2),\mathcal{C}_{di}^{\prime
}(3):d\in \mathbb{D}\right\} $ denote those sets of CPTs for a new entry which have been
elicited as different from any yet to appear in the library - i.e. for $%
j\in\{1,2,3\}$, with $i=1,2,\ldots,k$ indexing the models in the library, define 
\begin{equation*}
\mathcal{C}_{di}^{\prime }(j)\triangleq \mathcal{C}_{di}(j)\backslash \left(
\cup _{1\leq i^{\prime }<i}\mathcal{C}_{di^{\prime }}^{\prime }(j)\right)
\end{equation*}

Note that, for this construction to make sense, we have assumed we have
labelled the vertices in $V(\mathcal{G}_{i}), i=1,2,\ldots,k$, such
that vertices with the same index have the same meaning 
across different entries in the library - here across different criminal
plots. In the protocol described below, academics will need to craft the
naming of vertices so that these are as generic as possible so as to make
the association across different entries in the libraries as fluid as they can
be. We note that it is often necessary to revisit a generic naming of
vertices so that the meaning of the vertices continues to apply to all
entries (see e.g. \cite{Hatummodeltransfer}). We also assume that this will mean
that vertices, if they appear in two different graphs, will be ordered
compatibly with each other. Thirdly, in the protocol defined below, we will
assume that the graphs describing the criminal missions - here plots -
have been chosen by academics so that these will be causal in the senses we
have discussed above. This will mean that, for each category of crime, the
graphs of the progress of the criminal mission will be respected before and after any
intervention police might contemplate.

Because the libraries we construct have entries that describe similar
crimes, logic often demands, or it is at least plausible, to assume that the
dependence structures they express through the topology of the graphs within
a library are shared. We have also argued that some of their CPTs will also
be shared. It is therefore helpful the introduce a notation which can
reflect these commonalities over the $k$ models already located within the library. So let $\mathcal{G}_*\triangleq \cap _{i=1}^k
\mathcal{G}_{i}$ denote the graph with vertex set $V(\mathcal{G}
_*)\triangleq \cap _{i=1}^kV(\mathcal{G}_{i})$ with a directed edge $%
(v^{\prime },v)$ from $v^{\prime }$ to $v$ in the edge set $V(\mathcal{G}%
_*)$ if and only if the edge $(v^{\prime },v)$ lies in at least one of the edge sets $E(%
\mathcal{G}_{i})$, $i\in\{1,2,...,k\}$. Similarly, where $V_*\subseteq V(\mathcal{G}_*)$, let $\mathcal{C}_{d*}\triangleq \left\{ C_{dv*}:v\in V_*\right\}$ for some $d\in \mathbb{D}$ denote the set of those CPTs that are shared by all graphs in the library $\mathcal{G}_i, i \in\{1,2,\ldots,k\}$. Note that a necessary condition for $v\in V_*$ is that $v$ has the same
parents in each $\mathcal{G}_i, i\in\{1,2,\ldots,k\}$, so that all CPTs
have the same dimensions. This notation sets up a way to find these common CPTs in a large library to aid the construction of priors for a new library entry.

Before the construction of the library begins, the academic co-creators train
at least one of the police team so that they are able to elicit from
colleagues the probabilities that might be needed for CPTs whose values
must remain behind the firewall. This would typically encourage them to sign
up for one of several open courses in probabilistic elicitation methods. Some of these methods include variations of the Delphi method (see e.g. \cite{Rowe1999}), Cooke's classical method \cite{Cooke1991} and the IDEA protocol \cite{Hanea2017IDEA}, all of which involve asking groups of experts for their probabilistic judgements and evaluating these judgements over a number of stages. These tend to rely on the mathematical aggregation of experts' judgements as a consensus is not usually naturally reached among the experts. However, an expert panel formed of members of police teams who are accustomed to working together are more likely to reach a consensus about probabilistic judgements surrounding an ongoing criminal plot. Therefore, elicitation methods focused on behavioural aggregation may be favoured over the aforementioned methods. The main elicitation technique of this kind is the Sheffield Elicitation Framework (SHELF) \cite{Gosling2018} in which a facilitator guides the sharing of information and leads group discussions with the aim of reaching a consensus among experts. Far more detail on these elicitation methods can be seen in \cite{o2006uncertain,EFSAGuidance2014}. Members of the in-house police team now trained in such an elicitation method would then
receive more customised training via the academic co-creation team. Such
activities might involve the academic team engaging them in elicitations of
the CPTs within the first iteration of the academic library then donated
to police as a template of the model behind the firewall.

Denote the library built up by academics outside the firewall (police behind
the firewall) on the $t^{th}$iteration of development by $\mathbb{L}_{t}^{A}$ $\left( \mathbb{L}_{t}^{B}\right)$, and use the same labelling convention
for all entries and their pairs within these libraries. We write the existing entries in the libraries as $\left\{ M_{1},M_{2},\ldots ,M_{k-1}\right\} $, but these libraries may be empty. We are now able to describe a protocol for co-creating the
next entry $M_{k}$ into this library.

\subsubsection{Step 1 - Initial library based on open source information}

The academic team first set up the initial prototype version of their expanded
library $\mathbb{L}_{1}^{A}$ as follows:

\begin{enumerate}
\item Informed by previous studies undertaken when building $\left\{
M_{1}^{A},M_{2}^{A},\ldots ,M_{k-1}^{A}\right\} $ within the current
academic library $\mathbb{L}_{0}^{A}$, supplemented by other open source
information pertinent to $M_{k}^{A}$, and guided by police sharing their own
open source knowledge, academics choose a graph $\mathcal{G}_{k}$ representing the BN
of the next category and type of crime.\textit{\ }This stage typically involves
further engagement with experts and a deep dive into literature to discover
new information about the new entry to the library\textit{.}

\item The names of the vertices in $V(\mathcal{G}_{k}^{A})$ are made as
compatible as possible with the names given to vertices in $V(\mathcal{G}%
_{i}^{A})$, $i=1,2,\ldots ,k-1$. Note that this sometimes entails the
relabelling of the vertices in this set in light of the meaning of
vertices in the new vertex set $V(\mathcal{G}_{k}^{A})$. This is a delicate
process - see e.g. \cite{Hatummodeltransfer}, albeit in the very different
context of ecological modelling. This harmonisation step helps to maximise
the amount of local structure and probabilistic information that can be
shared across different BNs in the library and so helps minimise duplicating
effort establishing the next entry in the library.

\item We next begin populating the CPTs as these apply to the new entry of
the library which would be valid were police not to intervene. The academic
team first elicit from police the likely nature of the partition of the CPTs 
$\left\{ \mathcal{C}_{k}^{A}(1),\mathcal{C}_{k}^{A}(2),\mathcal{C}%
_{k}^{A}(3)\right\} $. They then need to elicit from the police team which
of these CPTs - because of their shared meaning with other models in the
library - already appear in the library, and which CPTs $\left\{ \mathcal{C}%
_{k}^{^{\prime }A}(1),\mathcal{C}_{k}^{^{\prime }A}(2),\mathcal{C}%
_{k}^{^{\prime }A}(3)\right\} $ are unique and hence need to be populated.

\item To populate $\mathcal{C}_{k}^{^{\prime }A}(1)$ and $\mathcal{C}%
_{k}^{^{\prime }A}(2)$, the academic team then proceed as they would in
non-secure settings. They first elicit the uncertain probabilities in $%
\mathcal{C}_{k}^{^{\prime }A}(1)$. They then refine these judgements based
on available data using Bayes Rule to construct open source posterior
tables. They repeat this process for CPTs in $\mathcal{C}%
_{k}^{A}(2)$. At this stage the academic team can use their skills to
identify and adapt statistical and AI methodologies, in particular to use
time series data extracted from particular past instances to calibrate and
efficiently estimate the parameters associated with these probability
tables.

\item Dummy entries are then chosen by the academic teams for those new CPTs
in $\mathcal{C}_{i}^{^{\prime }A}(3)$ that will become informed primarily
through secure in-house information, and labelled as such. We nevertheless
recommend that these dummy entries are chosen to appear as plausible as
possible to police and, where possible, calibrated against any available open
source case studies or outputs - at least as they might apply to one
category of suspect $\mathcal{S}$.

\item We note that different CPTs elicited here may need to be selected for
different types of suspect $\mathcal{S}$. Thus the MO and training of a
right-wing terrorist recruit might be very different from a terrorist
affiliated to IS which in turn may be very different from a suspect who is
acting completely autonomously. In our running examples, it will
usually be possible for police to reliably categorise any given triaged
suspect, although, in some plot libraries - like those designed to protect
against exfiltration attacks - such prior categorisation will be less
certain. In either case, when the library is designed to be applied to make
inferences about different categories of criminal, different
collections of CPTs will need to be constructed for each such category.
However, we note that most of the CPTs will be shared across different
categories, but those that \emph{do} differ help to formally discriminate the
possible type of suspect faced by police when this is uncertain.

\item Academics next need to populate their CPTs $\left\{ \mathcal{C}%
_{di}:d\neq d_{\phi }\in \mathbb{D}\right\} $associated with each potential
intervention they contemplate making. To do this, they repeat the elicitation
process described for the unintervened process in the 3 bullets above for
different categories of suspect $\mathcal{S}$. Superficially, this might look to be a
very large task. However, if the BN is well-chosen, whenever it can be
described as causal, as we have argued that
plot models can be \cite{Ramiahplot24}, this will typically only require the addition of a few
select CPTs for each potential intervention. This will be so, even if it
will be apparent to the criminal that such interventions have been made.

\item The models are then coded up as software. The code developed by the academic team will be much larger than the distilled code that will be delivered to the police team in order for the academic team to be able to explore various modelling choices that the police team do not need to do themselves, as well as to allow rigorous verification to be performed before the library is transferred to police. The academic team
check the plausibility of the outputs of this code and the faithfulness of
the code itself against synthetic cases. The academic team first simulate
use cases from their model. They then use open source data about real past
incidents, supplementing this with any synthetic data about records they
believe might have been observed but are now lost, performing the standard
statistical diagnostics normally used to check the performance of this BN.

\item As part of their development, the academic team will have embedded a
suite of estimation modules and diagnostic checks for the new model using
this data for police to later replicate. Examples illustrating this
process can be found in \cite{BunninSmith2021, Shenvi2023, BunninSmithJSM2021}.

\item The rationale behind both the chosen structures within the library,
the real CPT, any data used to calibrate these and the methodology to
accommodate them are all recorded for future appraisal by police. It is
vital to carefully provide the in-house team with a report carefully
documenting the rationale behind the choice of model for the new category
and to add this to any other such documentation as this applies to previous
models in the library. For an example of the description of the outputs of
such software and their embedded algorithms, see \cite{BunninSmith2021, Shenvi2023, BunninSmithJSM2021}.

\item Because these statistical models are all based on open source
information, they can be freely submitted to proper peer review and
criticism. The models, methodologies and applications can thus be properly
quality controlled to this point for later technical revision if this is
necessary at the earliest stage of the co-creation.

\item A handbook is created or modified for the new library. This includes
how the new well-documented transferred code works. It also demonstrates
the estimation, statistical diagnostics and dummy examples provided which in-house statisticians are then able to replicate for
the recent library entry. The handbook and a distillation of the code - both based solely on all
open source materials - is then delivered to the police team.
\end{enumerate}

\subsubsection{Step 2 - Police team create their first in-house library}

At this point, the police team will have received the latest
version of the academic library $\mathbb{L}_{1}^{A}$, including the model of the newest
crime entry for a selection of categories $\mathcal{S}$. They will also have received a distillation of the code used by the academic team in building and verifying their models. The code shared with the police team will be heavily simplified and will only feature critical components for police to replicate the chosen methodology of the academic team. They now need to use
this code and academic library to help develop their parallel library $\mathbb{L}_{1}^{B}$,
informed by their current library $\mathbb{L}_{0}^{B}$ of coded models
located behind the firewall, to add the new entry into their own library.
This will involve refining $M_{k}^{A}$ with the enhanced data available only
to them to produce a new entry $M_{k}^{B}$ for their own library. They will
have the capability to run any of the coarse models in the latest academic
library $\mathbb{L}_{1}^{A}$ as well as the more refined coded models in the
latest police library $\mathbb{L}_{1}^{B}$. We recommend they translate the
latest crime model $M_{k}^{A}$ in the following way:

\begin{enumerate}
\item Police take the delivered code and the enhanced library $\mathbb{L}%
_{1}^{A}$ and test that they can run and emulate the results provided in the
handbook of the delivered system. This ensures that the new library $\mathbb{%
L}_{1}^{A}$ has been successfully transferred.

\item The performance of $M_{k}^{A}$ within $\mathbb{L}_{1}^{A}$ is then
applied by both co-creation teams to any available securely emulated data carefully
constructed and delivered by the police team. This is done through calibration using secure information to provide the academic team shareable, informative outputs. This can be used to check whether or not the outputs of the
academic model look broadly plausible to police,
given their model is only informed by open source data. If this is not the
case\ then the academic team need to liaise with the police team to adjust $%
M_{k}^{A}$. However, within this co-creation step it is important for both
teams to bear in mind that this quality control step will only be as good as
the in-house emulated data sets.

\item Conditional on this emulation being verified, the mismatch is likely
due to either the misspecification of the elicited graph $\mathcal{G}%
_{k}^{A} $ or the inaccuracy of the academic guesses about the secure
CPTs. In the former case, the academic team will need to perform further
elicitation to resolve this issue as they would in contexts where there is
no security issue. In the latter case, the
police team will need to give hints
about how the priors within the open source model might be better calibrated
to reality, or, if this information is too sensitive, to acknowledge the
discrepancy and nevertheless retain the mismatching entry.

\item The next step is to translate the extended library $\mathbb{L}_{1}^{A}$
containing the new entry $M_{k}$, taking the current police library $\mathbb{%
L}_{0}^{B}$ behind the firewall and adding an adjusted version of this model
to form an initial construction of $\mathbb{L}_{1}^{B}$. Note that any non-empty extant police library $%
\mathbb{L}_{0}^{B}$ will typically contain a more refined suite of models
than the entries in the library $\mathbb{L}_{0}^{A}$ used by academics. In
particular, the CPTs in $\mathcal{C}_{i}^{B}(2)$ and $\mathcal{C}_{i}^{B}(3),i\in\{1,2,\ldots ,k-1\}$, may be much more accurate than their equivalents in $%
\mathbb{L}_{1}^{A}$. This in turn will mean that the CPTs for $M_{k}$
matched to other models in the library should give more reliable results
when applying $\mathbb{L}_{1}^{B}$ than $\mathbb{L}_{1}^{A}$. This will need
to be acknowledged within this translation step.

\item Police then adjust the pre-existing BNs within this initial construction of $\mathbb{L}_{1}^{B} $ - such as changes to the node names to more generic terms so that these
will be consistent with the revised library. They then make any necessary adjustments to the
topology of $\mathcal{G}_{k}^{A}$ to contain any structural information
known only to them. In the case of plot models, these additions are most likely those associated with intensity measures they might secretly use to inform them about whether or not various tasks are being undertaken by a
suspect. This is because academic teams are more likely to make erroneous guesses about the highly secure information police have that form the intensity structure of the plot model than they are about the task and phase structures of the plot model. Any new types of undisclosable measurements they might have
available that relate to tasks in $M_{k}$ but to no other entry in the
library will need to be represented.

\item Police have been trained to elicit in-house any prior probabilities
needed for the secure CPTs $\left\{ \mathcal{C}_{di}:d\in \mathbb{D}\right\} 
$. This is the most delicate part of the operation to manage. It is useful
for any in-house representative who has not been trained in probabilistic
elicitation and who will be needed to act as a facilitator to first attend
one of the currently available aforementioned probability elicitation programmes.
We have also found that the in-house technician can often benefit by more
bespoke training delivered by the expert academic team where they are part of
a mock elicitation as directly appropriate as possible. The academic team
will be ready to answer any generic questions the in-house facilitator might
have about this process. This will need to be repeated for all categories of
suspect-environment pair $\mathcal{S}$.

\item Police then populate the prior probabilities needed for the secure
CPTs elicited and facilitated by the trained in-house representative behind
closed doors. These CPTs will usually consist of analogues in $\mathcal{C}%
_{k}^{B}(3)$ to the CPTs in $\mathcal{C}_{k}^{A}(3)$ - where academic guesses of the topological structure of the network are accurate - as well as the CPTs associated
with any new intensity measures that the academic team did not include in their model based on open source information.

\item Data is then embedded by the in-house expert, emulating how they have
seen the academic team do this for CPTs whose expert judgements are not
sensitive and whose training data is open source. Because the mathematical
equations and supporting code for performing these tasks tends to be generic,
such information can usually flow freely between the two teams.

\item Police will now be able to use their adjusted code and algorithms to make
their improved predictions and inferences. They then emulate the statistical
estimation and diagnostics they have seen the academic team apply to
calibrate their models and check their plausibility against the totality of
data they have available to them. If through this process they discover
inadequacies in their model, they share these with the academic team - see Step 3 below.

\item Otherwise they write their own in-house handbook to the new code using
the handbook of the open source software as a template to support other in-house users and to prepare for the next library entry.
\end{enumerate}

\subsubsection{Step 3 - Police and academics toggle models until requisite}

It is unlikely that this first version of the in-house library $\mathbb{L}%
_{1}^{B}$ is totally calibrated and well-estimated. Our experience has been that
there will be latent problems identified by police concerning the
validity and plausibility of the inferences the code might generate.
Problems can often include the code not running at all or being extremely slow,
the outputs becoming absurd, and the inferences being too uncertain to provide any useful level of decision support. However, the fact that $\mathbb{L}%
_{1}^{B}$ was seeded by $\mathbb{L}_{1}^{A}$ means that the academic team are
improving their current library so that it is fit for purpose.

The next step is to help the in-house team produce code which appears at
least superficially to produce plausible outputs. Because the academics
understand well the functionality of the methodology, algorithms and code
that seeded it, they are much better informed than they might otherwise be to
help the in-house team address these problems at arm's length.

\begin{enumerate}
\item For any identified problems about the outputs of their internal code, police feed back carefully sanitised points of concern about the code or outputs of their in-house models associated
with $M_{k}^{B}$ to academics. Examples might be "$M_{k}^{B}$ seems to consistently underestimate
threats of people of type x", "The vast amount of data we have available
means that estimating various parameters of the model using the analogues in 
$\mathbb{L}_{1}^{A}$ simply does not seem to work.", "We have no data at all
that might inform x given y, and no obvious source of expert judgement
either.", and "For realistically informed secure use cases, the system seems to
learn far too slowly to provide operational decision support." just to give a few.

\item Vigorous conversations now take place between the two teams about the
problems communicated in a sufficiently generic way to keep all confidential
information behind the firewall. The types of alterations the academic team
might perform may include changing the topology of the graph, the parameters, the probabilities, and the estimation algorithms in $\mathbb{L}_{1}^{A}$ for police to copy in
modifying entries in $\mathbb{L}_{1}^{B}$. Examples of replies from academics
might be "It appears we might have missed within our description of the
underlying processes some of the ways that these threats might happen -
perhaps we can find these?", "Method y has proved to be a very efficient
feature extraction algorithm in analogous settings. We provide a version of
this that you might be able to embed in your code.", "Perhaps we could use
alternative sources that at least give us some idea about y?", "Do you have
effect data sets about past incidents whose outputs might allow these
explanatory parameters to be calibrated? If so we will show you how to use
this to set plausible values of these probabilities" and "We suspect that you
have set the signal-to-noise ratios too high - if you set it in this way
then..." among others.
\item This co-creation step continues to be iterated until the model is
requisite, i.e. there are no remaining
shortcomings the in-house team can find that make the outputs of their
in-house version appear inadequate \cite{Phillips1984,smithbook}.

\item The in-house team check the performance of their model on confidential
data sets about past criminals or current suspects. Any new issues like the
above are then shared securely with the academic team to help find
solutions. Through this interaction, the teams make all necessary
adjustments and co-create the latest versions of the libraries we shall
denote by $\mathbb{L}_{2}^{A}$ and $\mathbb{L}_{2}^{B}.$

\item The in-house handbooks of these two libraries are rewritten
and annotated, templated on the previous handbooks and embedding all the new
structures and settings from this last interaction. In particular, this
should include the rationale for choosing its settings and any new synthetic
and real examples appropriate to each given library.

\item The academic library is now ready for the input of the next plot model
so that the co-creation development cycle described above can be repeated as
necessary.
\end{enumerate}

\subsubsection{Maintenance of the library as environments change}

Once an entry within the in-house library is functioning appropriately, it is
likely to need to be maintained. This is because criminal processes and the
ways they are most likely to be perpetrated evolve. However, because of the
way these models have been structured, once a graphical model has been set
up, it is relatively easy to maintain. So, for example, for typical plot
models, the phases a criminal needs to traverse and the speed they can do
this usually do not evolve quickly and these task probabilities only change
at a moderate pace. The type of tell-tale signs measured by intensities do
change quickly, of course, both in light of technological refinements and
the evolving MO of various potential criminals. But these issues can be
addressed simply through regular in-house refreshing of the CPTs and
occasional small topological modifications of the BN, and police will have
in-house specialists to facilitate this. Again, because the academic team have
seeded the code, they can actively engage in this maintenance - providing new
inputs to any open source components of the code and methodological advice
about how to refresh all parts of the model.

Such maintenance is vital to perform regularly but is relatively resource
cheap to the police once the initial system is in place. Again, because of
the parallel development, much of the necessary updating can be delegated to
the academic team and those elements that need to be securely protected can
be routinely updated in-house, guided by the well-informed academic team.

\section{Uncertainty Handling in a Secure Library}

\subsection{Introduction}

In this section, we describe how to address uncertainty handling when
following this protocol - acknowledging that many of the data streams that
might inform this process are incomplete. We will focus mainly on those
aspects of uncertainty management that are central to this secure context.

We begin by outlining how three types of additional structural information
about plot models greatly simplify the population of the CPTs of the BN.
We then review methodologies translated from extant sources associated
with learning within the more general class of Integrating Decision Support
Systems (IDSSs), for which the 2TDBN we use here is a special case, when
such CPTs are uncertain. This enables us to use the sporadic data police
have available to both refine prior settings for a given instance and embed
real-time inference about a given suspect as their actions unfold. In
particular, we discuss the use of global priors which keep the process
description modular and greatly aids inference across the whole library.

We then discuss how a prior for a particular model can be set up, and describe how, once set up, a model drawn from a library of
plot models can be applied to a particular individual suspected of being
engaged in a particular plot featured in the library. We discuss how the Bayesian
paradigm can be used to formally embed various different types of incomplete
and indirectly informative data sources that might be available to the
police. We end by discussing how these methods can also be applied by police
to generate probabilities for the likely impacts of various interventions that
they might consider making.

\subsubsection{Three special structural features of BNs of plots}

There exists extra implicit structural information that simplifies the
specification of the CPTs of the 2TDBN of plot models. The first is the
explicit assumption of a pattern of zeroes within the transition matrix $%
W_{t}|W_{t-1}$ of the phases capturing police knowledge that it is
impossible for the criminal to transition from one phase to particular other phases
in a single time step. For example, by definition it will not be possible to
transition from phase $w_{0}$ to any other phase or, under the usual choice
of time index, for someone to transition from $w_{1}$ (being recruited)
to $w_{5}$ (travelling to the scene of the attack). This means that there are
far fewer probabilities needing to be elicited to specify the CPTs than for a more
saturated model (see Appendix A).

Second, for plot models, we assume that, in the active phase, the
distribution of tasks and their relationship structure can be represented
by a given BN independent of the phase of the plot the suspect is in \cite{Ramiahplot24}. Of course, the conditional distributions of the
tasks for each active phase, as well as for the inactive phase, will be distinct. Indeed these differences will help determine whether the suspect
is engaged in the plot and, if so, which phase they are likely to have
reached. This therefore justifies police focusing on particular
critical subsets of tasks to discriminate the current phase of the attack.
We give an explicit representation of this in the next section.

Finally, task sets define those tasks whose CPT entries for a given active phase are
different from those elicited conditional on the inactive phase $w_{0}$. For a given phase of a plot, many tasks are not critical, or indeed useful, for the transition of a suspect to a subsequent phase. Therefore,
many task CPTs for each active phase are simply duplicates of the task CPTs appearing for the inactive phase $w_{0}$, greatly simplifying the elicitation and inference task. See 
\cite{Ramiahplot24} for a more detailed discussion of this.

\subsection{Setting Up Uncertain Priors within a Library}

\subsubsection{Introduction}

Within a library of plot models, it is necessary to set up prior
distributions for each particular incident, mainly through prior estimates of CPT entries. These will be informed by past
incidents, some of whose developments will be exchangeable with the
unfolding case. The interesting challenge is whether it is possible to apply
standard formal Bayesian methodology to the different
library entries in a modular way, and whether the distributions of
the CPTs of the model within the library can be updated so that, posterior
to learning from similar past incidents and other sources, the distributions
remain modular.

The surprising answer to this question is `yes'; the necessary theory
and conditions for this property were established long ago for the BN (see e.g. \cite{LauritzenSpiegelhalter1988, Cowell1999}) and in fact remain
true even when applied to models with other graphical structures (see e.g. \cite{leonelli2015bayesian, Leonelli2020}). These earlier developments were
designed for single models, but they transfer straightforwardly and
seamlessly on to inferences within libraries in ways we outline below.

In this section, we first review the conditions for when a single BN
can be updated such that the posterior distributions of each of its
contributing CPTs can be updated functionally independently. We then apply
this result to develop analogous results about libraries of such models
where different collections of models and CPTs are assumed exchangeable across
the library. For the purposes of this paper we will focus only on this
modular learning as this applies to the BN, although analogous results apply
to libraries of other types of graphical model too.

\subsubsection{Modular learning on a single BN}

The assumption that the matrices of prior entries across the different CPTs of a BN are mutually independent is
called \emph{global independence} \cite{Cowell1999,smithbook}. There
are always globally independent priors available for any BN because - unlike
some other graphical models - its CPT entries are functionally independent.
Although this assumption can be broken - for example, when a facilitator
believes that an expert from whom the conditional probabilities elicited
from different CPTs may exhibit the same biases - most routine choices of
prior which express only vague judgements would automatically satisfy this
condition. When, as in our running example, all the CPTs of the 2TDBNs in
our library of models are finite and discrete, one obvious choice is to
give all the rows of each CPT independent Dirichlet distributions or
independent logistic distributions. These would give two examples of
globally independent priors on the entries of the model.

This assumption is so useful as the property of global
independence continues to hold posterior to data being used to refine the
probabilistic expert judgements within each of the different CPTs of the BN.
There are two such types of data that preserve this property:

\begin{enumerate}
\item \emph{Ancestral data} from given past incidents. If, after such
events, the values of all the variables can be reconstructed up to a given
point, so that if data is available for a particular variable then it is
also available for all its parents, then global independence will automatically
hold a posteriori. This means we can simply perform a prior-to-posterior analysis on each of the CPTs - using relevant count data from
criminals advancing through the phases of their plots.

\item \emph{Designed random sampling data} from relevant observational
studies. Here we choose to sample from various CPTs in the model - assumed
to be invariant across incidents - stratifying the sample across the
different combinations of the levels within the CPTs. When considering a
library of plot models, we would recommend sampling from the task
CPTs conditioning on the inactive phase $w_{0}$. Note that such sampling
data will usually be open source being informed by actions of innocent
people and so shared across the libraries of both the academic and in-house
team. Some of the task dependence data associated with criminal
activities may only be available to the in-house team, however, and so can only
be accommodated in this way in-house.
\end{enumerate}

The reason why global independence continues to hold in such settings is
that, in both cases, the likelihood on the parameters in the different CPTs 
\emph{separates} (for a formal definition of this, see e.g. \cite{smithbook}).
It means, for example, that it is simple for the in-house team to update their
priors using the secure information available only to them in addition to
the open source data available to the academic team. Both can then use a
simple Dirichlet conjugate prior-to-posterior analysis whose relevant
formulae the academic team can relay across the firewall, as well as simple
exponential waiting times for CPTs when the underlying phase transition process is assumed to be a semi-Markov process \cite{Shenvi2024}. In this setting, the in-house model will mirror the model from the academic team, but will feature different values for its
hyperparameters. There is now vast literature and training courses
available which the academic team can use to enable police to do this.

On the other hand, when we need to formally update models with data that is not
ancestral or designed - often data suffering from missingness - typically global independence posterior to such
data no longer holds. Indeed inferences from such data can be
very subtle and the signals it can give often ambiguous. For some examples
of the sorts of ambiguity of interpretation that can occur in even the
simplest BNs exhibiting such non-ancestral missingness, see, for example, \cite{CroftSmith2003, Mond2003,Zwiernik2011,Zwiernik2012}. Note
from these papers in particular that when interior variables are \emph{systematically} missing then there may well exist completely different
explanations of the modelled system equally well-supported by the data - i.e. having
the same observed likelihood - no matter how large the sample size.
In such cases, it becomes imperative that strong prior information is
introduced into the model to discriminate between these explanations. In
particular, retaining approximate global independence without introducing
such strong prior information is hopeless. Even in less extreme cases, there
will usually be a necessary trade-off between retaining the clarity,
interpretability and in-house calculability of the resulting inference and
its completeness. This judgement over the prior information used can only be expected to be possibly made
by a well-equipped statistician or machine learner - usually only one of the
academic team members.

Formal fast inferential Bayesian methodologies have already been developed
and can help address missingness of various types. This technology has
already been successfully applied to a number of recent applications of less
sensitive domains such as food security \cite{Barons2022foodsecurity} and to energy
systems \cite{Volodina22}. The processes here are informed through
composites of streaming state space time series whose components only
inform the states describing various subsets of these states but not others. In this context, it is wise to try to ensure the inference is as
formal as it can be without compromising its interpretability to police. We
would therefore recommend that one pragmatic solution is to follow the steps
below:

\begin{enumerate}
\item Both teams accommodate all available open source data whose likelihood
separates into their shared CPTs in the way we describe above.

\item The academic team embed any further open source data that destroys
global independence but strongly informs the inference about some of the
probabilities that otherwise would be extremely uncertain to both teams, and
then updates this joint distribution numerically \cite{Barons2022foodsecurity} or
with appropriate partial inference \cite{Volodina22}. Then the
academic team approximate this posterior with a conjugate distribution - a
product Dirichlet (or a mixture) exhibiting global
independence across the CPTs with exponential (or more general) waiting times \cite{Shenvi2024}. The academic team need to ensure that such an
approximation is a close one in an appropriate sense or abort the
accommodation of such data. These distributions are then translated in the
library $\mathbb{L}_{B}$.

\item The in-house team now refine their in-house distributions of their
library of CPTs using the secure data available only to them whose
likelihood separates into their shared CPTs, replacing the independent
posterior densities provided in $\mathbb{L}_{A}$ to provide new
distributions for these conditional probabilities in $\mathbb{L}_{B}$.

\item All other unused in-house and open source data are then simply used to
inform and adjust the setting of priors only for future incidents as police
deem appropriate.
\end{enumerate}

\subsection{Learning CPTs from Past Incidents and Across Different Graphs}

To construct a library, it is important that, as far as possible, the
distribution of the probabilities within the in-house library are supported
by a formal Bayesian analysis. Here the type of use to which the library
will be put is critical when designing the level of complexity of the
supporting formal estimation of its CPTs. Sometimes libraries are built for
the primary purpose of defining defences for a multitude of future
incidents. One example of such a library of BNs is the support of networks of
infrastructures to the increasing threats of flooding \cite{CREDO2022TR3}. For such purposes where we need to predict a population of
incidents, the dependence structure between estimated CPTs across
different instances within the same class has a big impact on the assessment
of the efficacy of different policy interventions. For example, the
consequences when these future unfoldings across many different incidents
are all identical will be very different to the case when each future
instance is an independent realisation from the same conditional
distribution.

However, when the primary use of the library is for real-time decision
support and so to guide interventions that apply only to the next incident,
inference about this dependence is not relevant. For this use - the
one we focus on in this paper - the in-house team need to estimate only the
expectations of the probabilities in the CPTs. This can often be done by
quite na\"ive estimation and so is relatively easy for the academic team to guide remotely.

More explicitly, suppose the in-house team need to defend the next unfolding
incident. Then inference simply corresponds to the
implementation of standard propagation algorithms for a BN - here the
relevant 2TDBN. We describe this more explicitly below. Note
that the population of each of the CPTs in the
library can be performed offline for each category of
incident and suspect the police might encounter, and this is then plugged into the
model of the current incident. The precise nature of this estimation -
respecting its types and patterns of missingness - can be applied to each CPT
in turn using various techniques discussed elsewhere in this paper.

\subsection{Learning within an Ongoing Incident with a Known Suspected
Perpetrator}

Thus suppose police are faced with an unfolding incident and a suspect is
then monitored in real time. We would recommend they
follow the procedure outlined below:

\begin{enumerate}
\item Police match as closely as they can the category of incident they face
to one they have catalogued within their library. Differentials between models in the library will often include the likely affiliation of the suspect, their broad intent and their
likely expression of attack - here the type of plot they are likely to
execute (for example, a bomb attack or a vehicle attack). We assume that their library is rich enough to contain a plot model whose structure sufficiently matches
the suspected incident.

\item Police then need to construct an appropriate set of CPTs that
match the current incident. Depending on the maturity of the library, and
depending on the category of incident and suspect, there will be CPTs of this
2TDBN populated with benchmark probabilities - whose expectations have been
estimated from an archive of previous incidents and domain knowledge. Police embed these into the model whenever these benchmark generic
expectations are available. Here we have assumed this is the case for most of the
CPTs. Note that after an incident has occurred, and particularly if this goes
to court, then values of variables associated with tasks and phases will
often become subsequently available. These can then be used to calibrate these
otherwise latent variables using the classifications mentioned above. This is especially useful for the matching of future incidents of the same or similar category to the current culminated plot, allowing a more accurate and calibrated translation of CPT entries from the developed library to ongoing cases.

\item Police then elicit the probabilities they need in-house to complete
the probability model of the suspected perpetrator and incident based on
private and covert information they have about the given suspect
that goes beyond the category found in the library best matching the current case.

\item Such secure information is then embedded by the in-house team to
sensitively adjust any entries in the CPTs to the current case, using the
techniques they have developed through teaching from the academic team and through training in elicitation methods.

\item Police then simply apply this model to the current unfolding
incident - embedding the specific routinely collected intensity
data as well as sporadic intelligence data to update predictions
concerning the efficacy of any interventions they contemplate making. How
this works has been documented elsewhere \cite{BunninSmith2021,BunninSmithJSM2021,Shenvi2023}.
\end{enumerate}

We note that the scope of the application of the 2TDBN software to support
decisions in this way depends on the existence of benchmark CPTs - estimated
as described above. Of course, standard Bayesian diagnostics would be used
in-house by police to determine whether or not there was information within
the unfolding incident to suggest that the choice of model was
inappropriate. If this is so, then police will need to disregard the outputs
of the support tool and reconsult the academic team.

Note that, within the Bayesian paradigm, if it is initially uncertain into which category a suspect falls, police can simply assign
a probability across these different categories - for example, over different types of attack or over different intents and capabilities of the perpetrator - and use the unfolding intensity data to discriminate between these
using standard Bayesian techniques for mixture modelling as directed by the
academic team. The only practical difficulty in setting this up is that
they would then need to follow the steps outlined above for each of the plausible candidate models.

Of course, once the incident has concluded, further data relevant to the incident that was not previously accessible to police will often become available in the form of court hearings and incident reports. Further resources may be employed to secure evidence in the prosecution proceedings, providing more data to police than was available at the point of utilising the model for decision support. Police can use such data to
update their probability distributions on the CPTs. Interestingly, if the
suspect eventually turns out to be innocent, then data collected by police
tagged to this person cannot be retained. However, they can first use this information to better inform the associated
CPTs in-house. Notice that this would help mitigate the biases that might
occur within open source data because of the missingness of records about
how innocent people might behave. Through adjusting the utilised model in light of these new data sources, it ensures that, when the model returns to the library for this category of suspect and environment, it is as accurate and reliable as feasibly possible so that, if it is called upon as a best match for a future case, it is best equipped to mitigate the challenges of the missingness of current data for that future case.

\subsection{Learning through Intensities}

Experience has taught us that much of the information that is to be kept secure concerns
what police can covertly monitor, as well as the reliability of this data, to filter out the tasks
the suspect might be engaged in. This impacts the application of the
protocol above in two ways:

\begin{enumerate}
\item The guesses the academic team make about these aspects of the model
will often be unreliable and so the in-house team will need to override a
significant proportion of these variables, particularly updating numerous entries of the
corresponding CPTs.

\item The available streaming information concerning one incident and
suspect can be very different to that available for another incident and suspect. Therefore, when populating the BN for a
particular incident, this information will often need to be bespoke by default
and hence entered by police in-house.
\end{enumerate}

To address this issue, we recommend that the academic team simply guess the
\emph{types} of tell-tale signs about such variables and provide a
variety of probability settings for the corresponding CPT entries for the
in-house team - reflecting the quality of the information from the
hypothesised intensity measures - to match their scenario. These generic
suggestions are particularly useful to help the in-house team become aware of the implications of
inferences they might make associated with past cases extended into
further hypothetical cases in different settings via particular learning algorithms.

A key component of the plot model is that many model parameters, especially those specific to the suspect $\mathcal{S}$, concern the speed at which the suspect can complete tasks and navigate phases of the plot. Experience has suggested that misspecification of
these parameters is not too critical provided these values are in the
right ball park. Parameters for which misspecification is the least critical are often furthest in topological
distance from the phase nodes. It can therefore be
proved that these probabilities have less effect than those in other CPTs in the
BN. See \cite{AlbrechtNicholson2014, Leonelli2024} for detailed discussion about this issue.

It might superficially appear that such a process would be difficult to
implement in real time. We end this section by using a hypothetical instance
of a particular vehicle attack of the type described above to illustrate why
this is proving not to be so.

\subsection{Co-Creation of Probability Models of a Vehicle Attack}

\subsubsection{Building on generic structures of plots into a BN}

The Bayesian model for the vehicle attack example was developed some time
ago. The basic generic structural model we originally developed has survived scrutiny and has provided a useful initial
template for many plots more generally. The phases of such attacks
were conveyed directly to us across the firewall and, after confirming these
through academic literature on the subject, we were able to hardwire the
corresponding non-zero states within the transition matrix $W_{t}|W_{t-1}$. It was surprisingly easy for the academic team to discover the strong impact of the category of
suspect (i.e. their background, including any sponsoring organisations, and their environment) on the speed of the development of a plot. The category of suspect was therefore clearly critical to the academic team
in determining transition probabilities for an ongoing incident, modelled surprisingly well with open source data.
The academic team then communicated their information through their
academic code. The main contribution of the in-house team was in providing summaries of their probabilistic beliefs that a particular category of suspect would
abort an attack, and that one of their triaged population of
suspects might be involved in a vehicle attack. The in-house team were then
encouraged to input values (which in any case they would need to review
on a case-by-case basis) specified directly by them.

Once we had elicited the task set as described above, we then needed to
specify the DBN of the tasks conditional on $w_{0}.$ In our initial coded
2TDBN, we made the heroic assumption that component tasks evolved
independently of each other. Although this assumption was extremely dubious,
it nevertheless provided a learning algorithm that worked very well for
predicting the phases of an attack. We would add that, in our current applications, we
are embedding open source information to estimate this sub-2TDBN directly.
This suffers no missingness associated with security issues because such
estimates can be based on benign observational data and extant sample
surveys, although some components do admittedly suffer from missingness issues
of a more conventional kind.

When a suspect is engaged in a phase of attack, the in-house team will
typically have good information - based on past experience and internal
intelligence about the particular suspect - about how they might act when
in a particular phase to move to a subsequent phase. However, even in this
case the academic team can provide probabilities based on plausible logical
guesses and open source past case studies to provide the in-house team a
reasonable benchmark from which to vary their own specification.

If police need to make predictions about what might happen were they to
intervene, they will need to revisit both the phase transition entries
and the task probabilities associated with the active states after any such
contemplated intervention. For example, they may choose to arrest the suspect
when the probability that the suspect has reached a sufficiently advanced phase of the attack is high enough to secure such an arrest.

Training for the in-house team to embed the additional CPTs needed for such
an embellishment is actually straightforward to provide, given that the
academic team ensure that the 2TDBN is \emph{adversarially causal}. This
is a delicate matter for the
academic team beyond the scope of this paper. For a detailed recent
discussion about how the academic team might do this in this context, see \cite{Ramiah2024, Ramiahplot24}. However, once the
underlying graphical framework has been confirmed as causal for the academic
model, then this property will translate seamlessly on to the in-house
replicate. So, with regard to missingness and firewall estimation, this
embellishment introduces no new challenges once the graphical framework is
in place. The intensity measures associated with each of these tasks can be
guessed by the academic team as discussed above.

\section{Discussion}

Having now developed this protocol, our academic team are currently populating
a number of different libraries of the type we have described above with
various in-house teams. We will report our findings later in this developmental process. However, even at this early stage we are able to
report some recent advances and challenges encountered when applying this
graphically enhanced technological transfer. We further aim to thoroughly apply a more diverse range of elicitation methods as discussed in Section \ref{NotationSetting}. We continue to explore new ways of faithfully eliciting prior judgements when information flows are disrupted between academic and in-house teams by security concerns of the type addressed here.


The BN is a powerful structural model, but it is not the only one, and, for
many classes of criminal activity, it is not the most efficient.
Even for plots (see e.g. \cite{Ramiahplot24}), we have argued for using
as a template not a 2TDBN but rather a customised hybrid graph with
different semantics. This is able to depict directly, and so more elegantly
and transparently, much more of the acknowledged structural information
directly. For example, through such constructions we can directly represent
the task sets as well as impossible phase transitions. The protocols above - expressed for the more generic family of 2TDBN\ - translate directly to such
applications. This is especially important for the study of models where,
unlike for the plot, the structural dynamics are poorly, rather than
inefficiently, expressed by a DBN, such as bot or incursion attacks \cite{Ramiah2024}. What is lost then of course is the valuable resource of
already extant suites of accessible quality-controlled software that
supports BN-based inference. So the choice of an appropriate structural
framework must be made on a case-by-case basis. But because police are in
any case constructing their own bespoke code behind a firewall, it is
often helpful to base these libraries around more customised graphs.

Clearly as the libraries of 2TDBNs - or other bespoke graphs - are built up, much more sophisticated methodologies can be developed to estimate the necessary
CPTs. Recent vigorous developments within adversarial risk
analysis \cite{Banks2015book,RiosInsua2016,RiosInsua2023} where causal hypotheses are plausible - as they are for plot models - are being harvested for use within this domain \cite{Ramiah2024,Ramiahplot24}. However, we have only just scratched the surface of the
potential of this symbiosis, and much fruitful work remains to be done to
exploit this extant technology and theory - in particular, to support,
through the types of co-creation described above, the in-house libraries of
crime incidents needed by the police.

One point we hope to have illustrated above is how - by employing a
subjective Bayesian approach - sparse, contaminated and sporadic data can
nevertheless be combined with informed expert judgements to build in-house
dynamic probability models for use by police in their pursuit of criminals
who they suspect are intent on extreme violence against the general
public. It is clear that terrorism intervention is not the only domain of application for which the protocols we describe in this paper are of significant benefit; these protocols could also be applied to asset, health and national infrastructure models, among many others. Within such contexts, missingness stemming from the high security of personal and national data is an inherent feature of such
analyses, but this is currently overcome by embedding strong and diverse structured expert judgements within such co-creation protocols.

\section*{Acknowledgements}
The authors would like to acknowledge the contributions of Professor Rob Procter and Dr Oliver Bunnin for their involvement in academic teams in the application and development of the protocols described in this paper, as well as the helpful comments of Dr Dave Hastie in supporting early versions of this paper and providing insights about software engineering for these modelling applications.

\appendix
\section{Structure of the Phase Transition Matrix}
The explicit form of the transition matrix between phases and its
parameterisation - which is retained even after intervention - is given below.
For the first phase, which is absorbing, we have that, for all $t\geq1$:

\begin{equation*}
p_{t}^{W}(w_{jt}|w_{0t-1})=\left\{ 
\begin{array}{cc}
1 & j=0 \\ 
0 & j\neq0
\end{array}%
\right.
\end{equation*}

while for $i=1,2,\ldots ,m$ and $t\geq1$:

\begin{equation*}
p_{t}^{W}(w_{jt}|w_{it-1})=\left\{ 
\begin{array}{cc}
q_{ti}^{\prime } & j=0 \\ 
(1-q_{ti}^{\prime })q_{ti} & 1\leq i=j\leq m \\ 
(1-q_{ti})(1-q_{ti}^{\prime })p_{tij}^{W} & 1\leq i\neq j\leq m,j\in E_{i}
\\ 
0 & 1\leq i\neq j\leq m,j\notin E_{i}%
\end{array}%
\right.
\end{equation*}
Here $q_{ti}^{\prime }$ denotes the probability that
the agent will abort the plot, and $q_{ti}$ the probability that the agent will remain in the same phase, both when in state $i$ and transitioning from time $t-1$ to time $t$. The probability $p_{tij}^{W}$
denotes the probability the next phase will be $w_{j}$ when in phase $w_{i}$
at time $t-1$ given the agent leaves their current phase and does not abort the plot.
$E_{i}$ are the states that can be reached from phase $w_{i}$ in one
time step. Note that when $E_{i}=\left\{ w_{j^{\ast }}\right\} $, so that
a criminal can only transition from phase $i$ to a single phase $w_{j^{\ast
}}$, then by definition $p_{tij^{\ast }}^{W}=1$ and so does not need to be
elicited.

\end{document}